\documentclass[11pt]{article}
\usepackage{amssymb,amsfonts,amsmath,latexsym,multirow,bbm}
\usepackage{color,float,fancyhdr}
\usepackage{graphicx,subfigure}
\usepackage[latin1]{inputenc}
\usepackage{pstricks,pst-node,pst-text,pst-3d}
\usepackage{algorithm,algorithmic}
\usepackage{cite}

\usepackage{parskip}

\usepackage{blindtext}
\usepackage{sectsty}
\usepackage[scaled=.60]{beramono}
\usepackage{courier}
\usepackage{parskip}
\usepackage{etoolbox}
\usepackage{subfig}

\usepackage[font={small,it}]{caption}

\allowdisplaybreaks 
\newcounter{subeqn} %

\definecolor{myred}{rgb}{0.5,0,0}
\definecolor{myblue}{rgb}{0,0,0.75}
\definecolor{mygreen}{rgb}{0,0.5,0}

\newtheorem{rem}{\sc Remark} [section]

\def\So{\mathcal{S}}
\newcommand{\R}{\mathbb{R}}
\newcommand{\N}{\mathbb{N}}


\sectionfont{\fontsize{12}{15}\selectfont}
\subsectionfont{\fontsize{12}{15}\selectfont}

\numberwithin{equation}{section}

\title{A self-calibrating method  for  heavy tailed data modelling.  Application in neuroscience and finance}

\author{Nehla Debbabi \thanks{ESPRIT School of Engineering, Tunis, Tunisia, \, E-mail: nehla.debbabi@esprit.tn}\, \thanks{Universit\'e de Reims Champagne Ardenne, CReSTIC Lab., France,\,  E-mail: mamadou.mboup@univ-reims.fr} \;, Marie Kratz \thanks{ESSEC Business School, CREAR risk research center; \, E-mail: kratz@essec.edu }\;, Mamadou Mboup \footnotemark[2]}

\date{December 20, 2017}

\begin{document}

\maketitle

\begin{abstract}
\noindent Modelling non-homogeneous and multi-component data is a problem that challenges scientific researchers in several fields. In general, it is not possible to find a simple and closed form probabilistic model to describe such data. That is why one often resorts to  non-parametric approaches.
However, when the multiple components are separable, parametric modelling becomes again tractable.
In this study, we propose a self-calibrating method to model multi-component data that exhibit heavy tails. We introduce a three-component hybrid distribution: a Gaussian distribution is linked to a Generalized Pareto one via an exponential distribution that bridges the gap between mean and  tail behaviors. An unsupervised algorithm is then developed for estimating  the parameters of this model. We study analytically and numerically its convergence. The effectiveness of the self-calibrating method is tested on simulated data, before applying it to real data from neuroscience and finance, respectively.  A comparison with other standard Extreme Value Theory approaches confirms the relevance and the practical advantage of this new method.
\end{abstract}


{\it 2010 AMS classification:} 60G70; 62E20; 62F35; 62P05; 62P10; 65D15; 68W40

\section{INTRODUCTION}

\vspace{-2ex}

Modelling non-homogeneous and multi-component data is a problem that challenges scientific researchers in several fields, as e.g. in climatology, finance and insurance, meteorology, and neuroscience (see e.g \cite{Sermpezis2015,Rangaswamy2004,Chao2012,Kollu2012,Rossi1984,Mandava2011,Debbabi2015a}).
In general, it is not possible to find a simple and closed form probabilistic model to describe such data. That is why one often resorts to  non-parametric approaches, such as e.g. kernel  density estimation ones (see e.g. \cite{kde,Parzen1962}) or non-parametric Bayesian methods (see e.g. \cite{BNPWalker99}), just to name a few.
However, when the multiple components are separable, parametric modelling becomes again tractable. Several hybrid models have been proposed in such context, combining two or more densities (see e.g. \cite{Frigessi2002,Czeledin2003,Behrens2004,Tancredi2006,Carreau2009,Mandava2011,Macdonald2011,Chao2012,Nadarajah2014,kratz2014}).

In this study, we tackle the general problem in a specific case, when data exhibit heavy tails. Extreme behaviors that are described by heavy tail modelling, can be observed  for a large number of phenomena, natural (from the big Dutch floods of 1953 to the earthquake of 2016 in Italy), financial (e.g. the sub-prime crisis in North America or the Sovereign debt crisis in Europe), medical (e.g. the  avian influenza), technological (e.g. Fukushima), or others.
One mathematical field, Extreme Value Theory (EVT), which started with \cite{Fisher_Tippett1928}, is totally devoted to the analysis and modelling of the extremes (see e.g. \cite{Embrechts1997,deHaan2006,Leadbetter1983,Resnick2007},  for general references).
Studies on extremes were developed in many fields, as, for instance, in financial markets and actuarial mathematics (see e.g. \cite{Feldman1998, Mcneil2005}), in epidemiology (see e.g. \cite{Guillou2014} for the first introduction of EVT in health surveillance), in signal processing (see e.g. \cite{Fine1962,Chellappa2010} when considering the general problem of false-alarm probability determination,  \cite{MBOUP12} for the spike detection in neural signals in biomedicine,  \cite{Milstein1969} for  the detection of a binary signal in additive  noise  in telecommunication, or \cite{Worden2002} for the damage detection  in machine diagnostics).

Introducing EVT helps managing the many catastrophes that our society is facing with, unfortunately, an observed  increasing trend of occurrence of extreme events since the beginning of the 20th century (see \cite{Dacorogna2015}). Moreover EVT methods help improving the standard data processing by taking into account the tail information.

Whereas EVT focuses on how to study and model extremes using the information in the tail of the distribution only (which is the strength of this theory, even if sometimes also its weakness in practice as tail data are scarce by definition), it is also very useful to combine it with standard statistics developed for the main information given in the data. To extract the important information given by extremes and to  highlight as well the information contained  in the entire underlying distribution, it is natural to take into account the asymmetry of the data weight above a high threshold (tail) and around the mean. Various mixture models, classified as parametric, semiparametric and nonparametric, have been proposed so far to do it  (see e.g. \cite{Scarrott2012} for a review of some of them).
  In this context, we suggest a parametric mixture  model to develop a self-calibrating  method for heavy tailed data modelling. The choice of this class of unsupervised procedures is clearly to ease practical implementations (in particular when complexity burden and/or delay processing are critical), but also to enlarge its applicability  (in particular with no assumption on any dependence type of the data).  Indeed, the difficulty faced when applying standard methods of EVT as the Peaks Over Threshold (POT) approach (see \cite{Davison1990}), or any other method to estimate the tail index (e.g. the Hill estimator (\cite{Hill}), to mention the most well-known 
 ; see also e.g. \cite{Longin2016}, chapters 5, 10 and 12), is that they are graphical ad hoc approaches, and concern often i.i.d. (heavy tailed) samples.

This self-calibrating method may be seen as two-folds: when (i) looking for a full modelling for non-homogeneous, multi-component and heavy tailed data, (ii) focusing on the tail and evaluating in an unsupervised way the high threshold over which the tail will be modeled; it might then constitute an alternative EVT method to standard ones as e.g. the POT approach.

We assume continuous (smooth transitions) and, with no loss of generality, right heavy tailed data (a similar treatment being possible on the left tail; see \cite{Debbabi2015a}).
How many components of the hybrid model to consider  and how to choose them? Since we are interested in fitting the whole distribution underlying heavy tailed data, the idea is to consider both the mean and tail behaviors, and to use limit theorems for each one (as suggested and developed analytically in \cite{kratz2014}), in order to make the model as general as possible. Therefore, we introduce a Gaussian distribution for the mean behavior, justified by the Central Limit Theorem (CLT), and a GPD for the tail, as the  Pickands  theorem (see \cite{Pickands}) tells us that the tail of the distribution may be evaluated through a GPD above a high threshold. Unlike existing two-component mixture models (see e.g. \cite{Frigessi2002,Behrens2004,Carreau2009,Debbabi2015b}), we bridge the gap between mean and  asymptotic behaviors by inserting  an exponential distribution.
A different weight is assigned to each component in order to have a better handling of the extremes.  The resulting three-component hybrid model is called G-E-GPD model.
Note that the GPD is the fixed component of this heavy tailed model, but the two other components could be chosen differently, depending on the data (and even reduced to one component, as in \cite{Debbabi2014a}).
Indeed, a specific treatment could be done to fit the exact distribution of the mean behavior for which we  have much data,  if one would like to avoid the use of the limiting normal distribution.  For instance, when having skewed distribution near the mean, the normal distribution could be replaced by a lognormal (as  generally done in insurance when fitting claims; see also e.g. \cite{Czeledin2003}). It would not change the idea of the self-calibrating method and of its algorithm.
Concerning the number of components, we point out that the model needs at least two-component, including the GPD, for the method to be workable. Indeed, the threshold over which the GPD is fitted (that we call the tail threshold), is determined in the algorithm as the junction point between the GPD and another distribution. It means some information before the tail threshold is required, contrary to standard EVT approaches.
Note the role of the intermediate distribution (here an exponential), used as a leverage to give full meaning of tail threshold to the last junction point between the GPD and its neighbour (this intermediate distribution).
The distance between two successive junction points will automatically tend to 0 if one component is not needed.

An iterative unsupervised algorithm is developed for estimating the parameters of the three-component hybrid model. It is based on the rule of estimating the model parameters in two steps, alternately and  iteratively, to avoid doing it at once, which might lead to substantially biased estimates.  It starts by enforcing the continuity and the differentiability of the three components at the two junction points, then proceeds in an iterative way to determine successive thresholds and parameters of the involved distributions. It provides a judicious weighting of the three distributions as well as a good location for the junction points or thresholds, especially for the tail threshold that emphasizes the presence of extremes. This algorithm is  based, for each iteration, on the resolution of numerical optimization problems in least squares sense, using the Levenberg Marquardt (LM) algorithm (e.g. \cite{Levenberg,Marquardt}). We study its convergence analytically and numerically.

The performance of this algorithmic method is studied in terms of goodness-of-fit on simulated data from G-E-GPD Monte-Carlo (MC) simulations.  Given the very good performance, we apply the method on real data, considering neural data and the S\&P500 log-returns. A comparison with other standard EVT approaches is also given.

The paper is organized as follows. In section \ref{sec:Model} we introduce our hybrid model. The method and its unsupervised iterative algorithm are developed in Section \ref{sec:Algo}.  Simulation results are presented in Section \ref{sec:Simul}, and applications of the method on real data in Section \ref{sec:Applic}; results are discussed and compared with those obtained via standard EVT methods in both sections. Conclusions follow in the last  section. The convergence of the algorithm and its robustness are discussed in Appendix \ref{sec:App1}.


\section{HYBRID THREE-COMPONENT MODEL}
\label{sec:Model}


 We consider a piecewise model where each component represents a different behavior of the data, which might be heterogeneous or not. We assume that the data admit a continuous (non-degenerate) distribution, and accordingly, introduce a general hybrid probability density function (pdf), with some smoothness constraints (\emph{i.e.} such that the transitions between components are smooth).
The hybrid model we propose, links three different distributions to each other at two junction points, denoted  by $u_1$ and $u_2$: a Gaussian distribution to model the mean behavior of the data,
a GPD to represent the tail, and an exponential distribution to bridge the gap between these two behaviors. This  model, denoted by G-E-GPD (Gaussian-Exponential-Generalized Pareto Distribution), is characterized by its pdf $h$ expressed as:
$$\displaystyle h(x;\boldsymbol{\theta})=\left\{
               \begin{array}{ll}
                \gamma_1\, f(x;\mu,\sigma) & \hbox{if}\;\;\; x\leq u_1, \\
                 \gamma_2 \, e(x;\lambda) & \hbox{if}\;\;\; u_1\leq x \leq u_2,\\
                 \gamma_3\, g(x-u_2;\xi,\beta) & \hbox{if}\;\;\;  x \geq u_2.\\
               \end{array}
             \right.$$
 The different parameters are gathered in a vector denoted by
$\boldsymbol{\theta}$ and are described hereafter. To begin, $\gamma_i$, $i\in\{1,2,3\}$,
stand for the weights associated to each component. The parameters $\mu\in\R$, and $\sigma\in\R_+^*=\R_+\backslash\{0\}$ represent,  respectively, the mean and the standard deviation  of the Gaussian pdf $f$ given by:
$\displaystyle f(x;\mu,\sigma)=\exp\{-(x-\mu)^2/2\sigma^2\}/\sigma\sqrt{2\pi}$, $\forall x\in\R.$
The parameters $\xi\in\R$ and $\beta\in\R_+^*$ denote,  respectively, the tail index and the shape parameter of the GPD pdf $g$, defined by:
$$
g(x;\xi,\beta)=\left\{
               \begin{array}{ll}
              \displaystyle   (1+\xi\,x/\beta)^{-1-\frac{1}{\xi}}/\beta & \hbox{if}\;\;\; \xi\neq 0, \\
                \displaystyle  e^{-x/\beta} /\beta & \hbox{if}\;\;\; \xi=0,
               \end{array}
             \right.\;\forall x\in \mathcal{D}(\xi,\beta):=\left\{
               \begin{array}{ll}
               [0,\infty) & \hbox{if}\;\;\; \xi\geq 0, \\

                 [0,-\beta/\xi] & \hbox{if}\;\;\; \xi<0.
               \end{array}
             \right.
$$
Finally,  $\lambda\in\R_+^*$ indicates the intensity parameter of the exponential  pdf $e$ defined by \linebreak[4] $\displaystyle e(x;\lambda)=\lambda \,e^{-\lambda x},\;\forall x\in\R_+.$\\
Imposing smooth transitions from one behavior to another, we constraint the hybrid pdf $h$ to be $\mathcal{C}^1$-regular. Note that combining this constraint and the heavy tailed data assumption will reduce the number of free
parameters, {\it i.e.} the size of $\boldsymbol{\theta}$. Let us present these assumptions.

\subsection*{Assumptions of the Model}

The first two assumptions are part of the construction of the G-E-GPD model.
\begin{itemize}
\item[(i)] First  we  assume, by construction,  that the data distribution  admits a pdf $h$.  This means that $h$ is non-negative and satisfies $\displaystyle \int_{\R} h(x;\boldsymbol{\theta})dx=1$, {\it i.e.}$$
 \gamma_1 F(u_1;\mu,\sigma)+\displaystyle \gamma_2\big(e^{-\lambda\,u_1}-e^{-\lambda\,u_2}\big)+\gamma_3=1,$$
where $F$ denotes the cumulative distribution function (cdf) of the normal distribution.
\item[(ii)] We  focus on heavy tailed data when $h$
  belongs to the  Fr\'{e}chet maximum domain of attraction ($\xi>0$);  therefore $\beta=\xi u_2$ (see \emph{e.g.} \cite{Embrechts1997}, p.159).
\end{itemize}

The main constraint is the smoothness of the pdf:
\begin{itemize}
\item[(iii)] The pdf $h$ is $\mathcal{C}^{1}$, {\it i.e.} is continuous and differentiable at the two junctions points $u_1$ and $u_2$.
\end{itemize}
Assumptions (i) to (iii) give rise to six equations relating all model parameters:
\begin{equation*}
\label{C3}
\left\{
  \begin{array}{ll}
   \beta=\xi \, u_2; & \gamma_1=\gamma_2 \,  \displaystyle \frac{e(u_1;\lambda)}{f(u_1;\mu,\sigma)}; \\
     \lambda=\frac{1+\xi}{\beta}; &
\gamma_2=\Big[ \xi \, e^{-\lambda\,u_2} +  \left(1+\lambda\,\displaystyle \frac{ F(u_1;\mu,\sigma)}{f(u_1;\mu,\sigma)} \right) e^{-\lambda\,u_1}\Big]^{-1};\\
  u_1=\mu+\lambda\sigma^2; & \gamma_3=\beta\,\gamma_2 \, e(u_2;\lambda).
  \end{array}
\right.
\end{equation*}
Consequently, the vector of the free parameters  is reduced to  $\boldsymbol{\theta} = [\mu, \sigma,  u_2, \xi]$.\\
It is then straightforward to deduce from $h$ the expression of the cdf and quantile function associated with the G-E-GPD model.
The G-E-GPD cdf, denoted $H$, is given by:
\begin{equation}
\label{cdf}
\displaystyle H(x;\boldsymbol{\theta})=\left\{
               \begin{array}{ll}
                \gamma_1 F(x;\mu,\sigma) & \hbox{if}\;\;\; x\leq u_1, \\
                 \gamma_1 F(u_1;\mu,\sigma)+\gamma_2 \big(e^{-\lambda\,u_1}-e^{-\lambda\,x}\big) & \hbox{if}\;\;\; u_1\leq x \leq u_2,\\
                 1-\gamma_3 \big(1+\xi(x-u_2)/\beta\big)^{-1/\xi} & \hbox{if}\;\;\;  x \geq u_2,\\
               \end{array}
             \right.
             \end{equation}
          and the corresponding quantile function by:
             $$\displaystyle H^{-1}\big(p;\boldsymbol{\theta}\big)=\left\{
               \begin{array}{ll}
                F^{-1}\left(\displaystyle p/\gamma_1;\, \mu,\sigma\right) & \hbox{if}\;\;\; p\leq p_1:= \gamma_1 F(u_1;\mu,\sigma), \\
                \lambda^{-1}\log\left(\displaystyle \frac{\gamma_2}{p_1-p + \gamma_2 e^{-\lambda\,u_1}}\right) & \hbox{if}\;\;\; p_1\leq p \leq p_2:=1-\gamma_3,\\
                \displaystyle \beta\left(\left[1-(p-p_2)/\gamma_3\right]^{-\xi}-1\right)/\xi \, +\, u_2 & \hbox{if}\;\;\;  p \geq p_2,\\
               \end{array}
             \right.$$
             where $F^{-1}$ denotes the  normal (mean $\mu$ and variance $\sigma^2$) quantile function.

The next question, due to the use of a parametric model, concerns the estimation of the parameters. To answer it, we develop an iterative algorithm for estimating $\boldsymbol{\theta}$, based on that built for two components in \cite{mlsp13_hal} and \cite{Debbabi2014a}. For each iteration, the LM algorithm (see \cite{Levenberg,Marquardt}) helps to solve numerically optimization problems in least squares sense. We describe our algorithm in the next section and prove its convergence analytically and numerically in Appendix \ref{sec:App1}.

\section{ITERATIVE ALGORITHM FOR HYBRID MODEL PARAMETERS ESTIMATION}
\label{sec:Algo}

Here we describe the iterative algorithm, which self-calibrates the G-E-GPD model, in particular the tail threshold above which a Fr\'echet distribution fits the extremes.   Its convergence is studied in Appendix \ref{sec:App1}; we prove analytically the existence of a stationary point, then numerically that the stationary point is attractive and unique.

This algorithm follows the same logic as the one developed for two components in \cite{mlsp13_hal} and \cite{Debbabi2014a}. For each iteration, the three-component algorithm  breaks down the problem of the  parameters vector $\boldsymbol{\theta}$ estimation into two nested subproblems; the parameters $\boldsymbol{p}=[\mu,\sigma,u_2]$ and tail index $\xi$ are estimated alternatively. Indeed, for each iteration, first we estimate the parameters vector $\boldsymbol{p}$ by minimizing the Squared Error (SE)  between the empirical cdf and the estimated one, when replacing $\xi$ by its estimate obtained in the previous iteration. Then, we estimate again $\xi$ by minimizing the SE between the empirical cdf and the estimated one, replacing this time $\boldsymbol{p}$ by its last estimate. Evidently, this procedure starts by fixing initial parameters and ends when a stop condition is satisfied.

First, we consider a sample $X=(X_i)_{1 \leq i\leq n}$ with  a G-E-GPD parent distribution, and denote by $\boldsymbol{x}=(x_i)_{1 \leq i\leq n}$ a given realization. For the rest of this work, $\widetilde{a}^{(0)}$ and $\widetilde{a}^{(k)}$ denote  the initialization and the estimate of the parameter $a$ at the $k^{th}$ iteration, respectively.

To run the iterative three-component algorithm, we  start initializing $\widetilde{p}^{(0)}=[\widetilde{\mu}^{(0)}, \widetilde{\sigma}^{(0)}, \widetilde{u}_2^{(0)}]$, instead of $\widetilde{\xi}^{(0)}$ since the only information we have about $\xi$  is its positivity.
To do so, we choose $\widetilde{\mu}^{(0)}$ as the mode of the data, and, according to the fact that about $16\%$ of Gaussian observations are below $\mu-\sigma$, we take $\widetilde{\sigma}^{(0)}=\widetilde{\mu}^{(0)}+q_{_{16\%}}$, where $q_{_{16\%}}$ represents the  empirical quantile of order $16\%$ associated to $H$. We also choose $\widetilde{u_2}^{(0)}$  as a quantile of high order $\rho$ (as we fit a GPD above $u_2$).
Then we use this initialization $\boldsymbol{\widetilde{p}^{(0)}}$ to determine $\widetilde{\xi}^{(0)}$, minimizing the SE between the hybrid cdf given $\boldsymbol{p}=\boldsymbol{\widetilde{p}^{(0)}}$ (fixed), and the empirical cdf $H_n$ associated to the $n$-sample $X=(X_i)_{1\leq i \leq n}$, defined, for all $t\in\R$, by
 $\displaystyle{H_n(t)=\sum_{i=1}^n \mathbbm{1}_{(X_i\leq t)}}/n$.
Note that we do not evaluate this SE on the observations $x_i$ only (as there might be only a few observations in the tail), but on a  generated  sequence of synthetic increasing  data  $\boldsymbol{y}=(y_j)_{1\leq j \leq m}$, of size $m$ (that may be different from  $n$),  with a logarithmic step, in order to increase the number of points above the tail threshold $u_2$. More precisely, for any $1\leq j\leq m$, $y_j$ is defined by:
 \begin{equation}
 \label{newdata}
 y_j=\underset{1\leq i \leq n}{\min(x_i)}+(\underset{1\leq i \leq n}{\max(x_i)}-\underset{1\leq i \leq n}{\min(x_i)})\log_{10}\left(1+\frac{9(j-1)}{m-1}\right).
 \end{equation}
Notice that the introduction of new points between the observations of $X$ has an impact on $H$ by evaluating it on more points, but not on the step function $H_n$.

Hence, $\widetilde{\xi}^{(0)}$ is now determined by solving the following minimization problem using the LM algorithm (see \cite{Levenberg,Marquardt}):
$$
\widetilde{\xi}^{(0)} \leftarrow \underset{\xi>0}{argmin}\left\| H(\boldsymbol{y};\,\boldsymbol{\theta}\mid{\boldsymbol{\widetilde{p}^{(0)}}})-H_n(\boldsymbol{y})\right\|_2^2,
$$
where  $\boldsymbol{\theta}\mid{\boldsymbol{\widetilde{p}^{(0)}}}$ represents  $\boldsymbol{\theta}$ for  $\boldsymbol{p}=\boldsymbol{\widetilde{p}^{(0)}}$ and $\|.\|_2$ denotes the Euclidean norm.

Once $\widetilde{\xi}^{(0)}$ is determined,  we can, thereafter, proceed iteratively. For all $k\geq 1$, the  $k^{th}$ iteration is splitted into two main minimization problems, which are solved alternatively, as described hereafter.\\
\textbf{\emph{Step $1$:}} Determination of $\widetilde{p}^{(k)}=[\widetilde{\mu}^{(k)}, \widetilde{\sigma}^{(k)}, \widetilde{u}_2^{(k)}]$, minimizing the SE between the hybrid cdf given $\widetilde{\xi}^{(k-1)}$, and the empirical one, as follows:
$$
\boldsymbol{\widetilde{p}^{(k)}} \leftarrow \underset{u_2 \in\R_+}{\underset{(\mu,\sigma) \in\R\times\R^*_+}{argmin}}\left\| H(\boldsymbol{y};\, \boldsymbol{\theta}\mid{\widetilde{\xi}^{(k-1)}})-H_n(\boldsymbol{y})\right\|_2^2
$$
where $\boldsymbol{\theta}\mid{\widetilde{\xi}^{(k-1)}}$ denotes $\boldsymbol{\theta}$ for $\xi=\widetilde{\xi}^{(k-1)}$ (fixed).\\
This minimization problem is as well numerically resolved using the LM algorithm.

\textbf{\emph{Step $2$:}} Determination of  $\widetilde{\xi}^{(k)}$,  minimizing the SE between the hybrid cdf given $\boldsymbol{\widetilde{p}^{(k)}}$, and the empirical one, {\it i.e.} by solving the following minimization problem via the LM algorithm:
$$
\widetilde{\xi}^{(k)} \leftarrow \underset{\xi>0}{argmin}\left\| H(\boldsymbol{y};\,\boldsymbol{\theta}\mid{\boldsymbol{\widetilde{p}^{(k)}}})-H_n(\boldsymbol{y})\right\|_2^2,
$$
where $\boldsymbol{\theta}\mid{\boldsymbol{\widetilde{p}^{(k)}}}$, represents $\boldsymbol{\theta}$ for $\boldsymbol{p}=\boldsymbol{\widetilde{p}^{(k)}}$ (fixed).

\textbf{\emph{Stop condition:}}
 The  algorithm iterates until  it satisfies the following stop condition:
$$
\left(\underset{\mbox{Condition\;C1}}{\underbrace{d\left(H(\boldsymbol{y};\boldsymbol{\theta}^{(k)}),H_n(\boldsymbol{y})\right)< \varepsilon }} \quad\mbox{and}\quad\underset{\mbox{Condition\; C2}}{\underbrace{d\left(H(\boldsymbol{y_{q_{_\alpha}}};\boldsymbol{\theta}^{(k)}),H_n(\boldsymbol{y_{q_{_\alpha}}})\right) <\varepsilon}}\right) \quad \text{or} \;\underset{\mbox{Condition\; C3}}{\underbrace{ k=k_{max}}}
$$

where $\varepsilon$ is a positive real that is small enough, $\boldsymbol{y_{q_{_\alpha}}}$ represents the observations above a fixed high quantile $q_{_{\alpha}}$ of arbitrary order $\alpha\ge 0.8$  associated with $H$ and $d(a,b)$ denotes the distance between $a$ and $b$, chosen in this study as the Mean Squared Error (MSE); it can be interpreted as the Cram\'{e}r-von-Mises test of goodness of fit.

To ensure a reliable fit of data not only for the main behavior but also for the tail, we force the algorithm to stop only when the MSE between the hybrid cdf and the empirical one is small enough, using on one hand all data (Condition C1), and on the other hand only extreme order statistics above $q_{_\alpha}$ (Condition C$_2$) (we chose $\alpha=0.8$ in our simulations and examples). Otherwise, the algorithm stops when a fixed number $k_{max}$ of iterations ($k_{max}=10^3$ in our simulations and examples) is reached (Condition C3).
\begin{rem}~
\begin{enumerate}
\item If focusing on the tail fit, an alternative to reduce the number of iterations of the algorithm, without playing with the value of $\varepsilon$, is to introduce the additional stop condition:\linebreak[4] $|\widetilde{\xi}^{(k)}-\widetilde{\xi}^{(k-1)}|<\varepsilon$.
 It stops the algorithm when the MSE value of Condition C1 or/and Condition C2 stops declining before reaching the chosen $\epsilon$.
 \item
This algorithm can be adapted to different hybrid models according to the  nature and the number of its components (if larger than 2), without any influence on the convergence study of the adapted algorithm.
\item
Although hybrid models considered in this study are assumed to belong to the Fr\'{e}chet maximum domain of attraction ({\it i.e.} for $\xi>0$), this algorithm can be extended to the case when the tail index of the GPD is free of constraints, as developed in \cite{Debbabi2014a} for a two-component model.

\item It would appear natural to estimate the hybrid model parameters via the Maximum Likelihood Estimation (MLE) method. However, in practice, the resort to this standard method becomes significantly complex when the number of free parameters is greater than 3, or when these latter are strongly linked in a non-linear way, as for our hybrid model. Estimating the 4 free parameters of the G-E-GPD model, all at once, using the MLE method becomes indeed challenging, mainly when choosing  initial parameters and determining the expression of the analytic gradient. Hence the choice of this alternative way to estimate the parameters, which is quite performant.
\end{enumerate}
\end{rem}


\section{SIMULATION RESULTS AND DISCUSSION}
\label{sec:Simul}


To study the performance of the algorithm to self-calibrate the  G-E-GPD model,  we build on MC simulations. To do so, we proceed in $4$ steps:

\begin{itemize}
  \item[Step $1$:] We consider  $N$ ($=100$ for this work) training sets $\displaystyle \{\boldsymbol{x^q}=(x_p^q)_{1\leq p\leq n}\}_{1\leq q \leq N}$ of length $n$, and $N$ test sets $\displaystyle \{\boldsymbol{y^q}=(y_p^q)_{1\leq p\leq l}\}_{1\leq q \leq N}$ of length $l$, with a G-E-GPD  parent distribution admitting  a fixed  parameters vector  $\boldsymbol{\theta}$.

  \item[Step $2$:] On each training set $\boldsymbol{x^q}$, $1\leq q \leq N$, we estimate $\boldsymbol{\theta}$, say $\widetilde{\boldsymbol{\theta}}^q=[\widetilde{\mu}^q, \widetilde{\sigma}^q,\widetilde{u_2}^q, \widetilde{\xi}^q]$, using the algorithm given in the previous section. We denote by $\widetilde{a}^q$ the estimate of the parameter $a$ relative to the $q^{th}$ training set.

  \item[Step $3$:]  We compute the empirical  mean and  variance of $\widetilde{a}^q$  over the $N$ training sets, namely \linebreak[4] $\displaystyle \tilde{a}=\sum_{q=1}^N \widetilde{a}^q/N$ and $\displaystyle \tilde{S}_N^{a}=\sum_{q=1}^N (\widetilde{a}^q-\widetilde{a})^2/(N-1)$, respectively. We  check the relevance of $\tilde{a}$  using two criteria:
        \begin{itemize}
          \item[(i)] The MSE expressed for any parameter $a$ as:
          $\displaystyle \mbox{MSE}_{a}=\sum_{q=1}^N (\widetilde{a}^q-a)^2/N.$ \linebreak[4]
          A small value of MSE highlights the reliability of parameters estimation using our algorithm.
          \item[(ii)]  Test on the mean (with unknown variance) :
$\displaystyle
\left|
\begin{array}{lcl}
H_0 &:& \widetilde{a}=a\\
H_1 &:& \widetilde{a}\neq a
\end{array}
\right.
$.\\
Since $N$ is large, we use a normal test (instead of a $t$-test) of size $\delta$, with a rejection region  of $H_0$ at risk $\delta\%$ described by
$\displaystyle \left(|T_{\tilde{a}}|>\Phi^{-1}(1-\delta/2)\right)$,
where the statistics $T_{\tilde{a}}$ is given by $\displaystyle  T_{\tilde{a}}=(\tilde{a}-a)/\sqrt{\tilde{S}_N^{a}},$ and $\Phi^{-1}(1-\delta/2)$ denotes the quantile of order $1-\delta/2$ of the standard normal distribution $\Phi$.
        \end{itemize}
  \item[Step $4$:] We compare the  hybrid pdf $h$ given $\boldsymbol{\theta}$ with the pdf $ \widetilde{h}$ estimated on each test set $\boldsymbol{y^q}$, given $\widetilde{\boldsymbol{\theta}}^q$. To do so, we compute the average of the log-likelihood ratio  $\mathcal{D}$ of $h(\boldsymbol{y^q};\boldsymbol{\theta})$ by $\widetilde{h}(\boldsymbol{y^q};\widetilde{\boldsymbol{\theta}}^q)$, over the $N$ simulations:
\begin{equation}\label{D}
  \mathcal{D}=\frac{1}{Nl}\sum_{q=1}^{N} \sum_{p=1}^{l} \log\left(\frac{h(y^q_p;\boldsymbol{\theta})}{\widetilde{h}(y^q_p;\widetilde{\boldsymbol{\theta}}^q)}\right).
  \end{equation}
 It is obvious that the smallest the value of $ \mathcal{D}$ is, the most trustworthy is the algorithm.
\end{itemize}

Several MC simulations have been performed varying $\boldsymbol{\theta}$ and $n$, to test the robustness of the algorithm. In this section, we illustrate the results we obtained with one example, and refer to  Appendix \ref{sec:App2} for more examples. In Table \ref{tab1}, we present the MC simulations results  when taking  $\boldsymbol{\theta}=[2,1,5,0.5]$,  $l=n$, $\delta=5\%$, $\alpha=0.8$ and $\rho=0.9$. Different values of $n$ have been considered to study its impact on the parameters estimation. The reliability of the three-component algorithm, in terms of goodness-of-fit, is pointed out through the two criteria described above and the  average of the log-likelihood ratio $ \mathcal{D}$. First, for each estimated parameter, we notice a small MSE whenever the data size is large enough, with a variance of order $1/n$, except for the threshold $u_2$ for which the variance is higher, varying in Table~\ref{tab1} as $n^{-1/3}$, $n^{-1/2}$, $n^{-3/5}$ for the three values of $n$, respectively. The variability is higher for the estimation of the threshold than for other parameters, well known phenomenon whatever is the method, and with no real impact on the estimation of the tail index, quite accurate as for the other parameters.

When varying the parameters (see Appendix \ref{sec:App2}), we observe similar results. The estimation of the tail index is quite robust, whatever its value (larger or smaller than 1), the size $n$, or the choice of thresholds $u_1$ and $u_2$.
Varying the thresholds $u_1$ or $u_2$ means changing the size of the subsamples used to estimate the parameters of the components of the hybrid cdf, hence some increase of the variability when the subsample size decreases, as expected. The variability is once again higher when estimating the threshold $u_2$.

It is worth noticing that the convergence with increasing $n$ is in $1/n$ for each parameter of the model, including the threshold (see Tables \ref{tab1}, \ref{tab11}, \ref{tab11bis}). It confirms the good quality of the performance. Indeed, although we look at the variability of each parameter  separately, one should not forget that we are fitting here a distribution, that is why we observe, as expected, a convergence in $1/n$. This is confirmed by the behavior of $ \mathcal{D}$ that highlights the accuracy of the parameters estimation, whatever $n$ (and also with a convergence in  $1/n$ with increasing $n$), hence the good self-calibration of the G-E-GPD hybrid model.

\begin{table}[H]
\caption{MC simulations results  for  $\boldsymbol{\theta}=[2,1,5,0.5]$, $\delta=5\%$, and $l=n \in \{10^3,10^4,10^5\}$.} \label{tab1}
\centering
\scalebox{1}{
\begin{tabular}{|c|c|c|c|c|c|}
\cline{4-6}
\multicolumn{3}{c|}{ } & $n=10^3$ & $n=10^4$ & $n=10^5$ \\ \hline
\multirow{8}{*}{\vspace{-7cm}\rotatebox{90}{Parameters}}&\multirow{4}{*}{$\mu=2$ } & $\widetilde{\mu}$ & $1.9981$ & $1.9994$ & $1.9994$\\  \cline{3-6}
&  & $\tilde{S}_N^{^\mu}$ & $6.87$ $10^{-3}$ & $8.91$ $10^{-4}$ & $7.68$ $10^{-5}$ \\  \cline{3-6}
&  & MSE$_{\mu}$ & $6.8$ $10^{-3}$ & $8.82$ $10^{-4}$ & $7.61$ $10^{-5}$ \\  \cline{3-6}
& & $T_{\widetilde{\mu},N}$ & $-0.0228$ & $ -0.0182$ & $ -0.0343$ \\  \cline{3-6}
\multirow{8}{*}{}&\multirow{4}{*}{$\sigma=1$ } & $p_{T_{\widetilde{\mu},N}}$ & $0.9818$ & $ 0.9854$ & $ 0.9726$ \\  \cline{2-6}
 & & $\widetilde{\sigma}$ & $1.0013$  & $ 1.0007$ & $0.9999$ \\   \cline{3-6}
 &  & $\tilde{S}_N^{^\sigma}$ & $4.73$ $10^{-3}$ & $4.88$ $10^{-4}$ & $5.23$ $10^{-5}$ \\  \cline{3-6}
 & & MSE$_{\sigma}$ & $4.69$ $10^{-3}$ & $4.83$ $10^{-4}$& $5.17$ $10^{-5}$ \\  \cline{3-6}
 & &  $T_{\widetilde{\sigma},N}$ & $0.0192$ & $0.033$ & $-0.0083$\\  \cline{3-6}
\multirow{8}{*}{}&\multirow{4}{*}{$u_2=5=q_{_{85.34\%}}$ } &  $p_{T_{\widetilde{\sigma},N}}$ & $0.9846$ & $ 0.9736$ & $0.9933$\\  \cline{2-6}
 & & $\widetilde{u_2}$ & $4.9904$ & $4.9896$  & $4.9964$\\   \cline{3-6}
 &  & $\tilde{S}_N^{^{u_2}}$ & $5.49$ $10^{-1}$ & $4.9$ $10^{-2}$ & $3.45$ $10^{-3}$ \\  \cline{3-6}
 & ($u_1=2.6=q_{_{53.88\%}}$) & MSE$_{u_2}$ & $5.43$ $10^{-1}$ & $4.86$ $10^{-2}$ & $3.43$ $10^{-3}$\\  \cline{3-6}
 & & $T_{\widetilde{u_2},N}$ & $ -0.0128$ & $-0.0466$ & $ -0.0599$\\  \cline{3-6}
\multirow{8}{*}{}&\multirow{4}{*}{$\xi=0.5$ } & $p_{T_{\widetilde{u_2},N}}$ & $ 0.9897$ & $0.9628$ & $0.9522$\\  \cline{2-6}
 & & $\widetilde{\xi}$& $0.4975$ & $0.5005$ & $5.0018$ \\   \cline{3-6}
  &  & $\tilde{S}_N^{^\xi}$ & $1.66$ $10^{-3}$ & $1.46$ $10^{-4}$ & $1.11$ $10^{-5}$ \\  \cline{3-6}
 & & MSE$_{\xi}$ & $1.65$ $10^{-3}$& $1.45$ $10^{-4}$& $1.1$ $10^{-5}$ \\  \cline{3-6}
 & & $T_{\widetilde{\xi},N}$ & $-0.061$ & $0.0459$& $ 0.0547$\\   \cline{3-6}
 & & $p_{T_{\widetilde{\xi},N}}$ & $0.9513$ & $0.9633$& $0.9563$\\   \hline
 \multicolumn{3}{|c|}{ Average  execution time (seconds)} &$3.83$  & $13.2$  & $245.68$ \\ \hline
  \multicolumn{3}{|c|}{ Average  iterations number} &$45$ &$48$ &$50$ \\ \hline
 \multicolumn{3}{|c|}{ $\mathcal{D}$} &$2.98$ $10^{- 3}$& $2.69$ $10^{- 4}$ & $2.98$ $10^{-5}$\\ \hline
\end{tabular}
}
\end{table}

We also resort to a statistical test as an additional criterion. For the $N$ training sets, we compute the test statistics denoted $T_{\tilde{a},N}$ and the corresponding $p$-value $p_{T_{\tilde{a},N}}=2(1-\Phi(|T_{\tilde{a},N}|))$, with respect to the parameter $a$, that we will compare to $\delta$. If this p-value is larger than $\delta$, we do not reject $H_0$.
For any $n \in \{10^3, 10^4, 10^5\}$ and for any  parameter $a \in \{\mu, \sigma, u_2, \xi\}$, we obtain $|T_{\tilde{a},N}|<\Phi^{-1}(0.975)=1.96$, and $\displaystyle p_{T_{\tilde{a},N}}>\delta=5\%$, which reveals a high acceptance (at $95\%$ level) of $H_0$ ($\tilde{a}=a$) \emph{i.e.} a very high level of similarity between the values obtained via the  algorithm and the fixed ones.

A remaining question, which might be the object of another paper, is the study of the rate of convergence for this algorithm. Here, to have an idea of how fast it works, we calculate
the average execution time and the average iterations number (the floor function) over the $N$ simulations. As shown in Tables \ref{tab1} and \ref{tab11}, they both  increase with  the data size, as expected.
 We notice that the average execution time  is small, even for $n=10^5$, indicating a fast convergence of the algorithm. It might be even reduced by converting our programs from the  R programming language to the C++ one.

 Besides the reliable estimation of the parameters, we show in Table \ref{tab2}, \emph{via} the MSE, that our algorithm enhances the GPD parameters estimation when compared with the Maximum Likelihood (ML) and the Probability  Weighted Moments (PWM) (see \cite{Hosking1987}) methods, performed when fixing the threshold at the value given by the algorithm, for fair comparison.

\begin{table}[H]
\renewcommand{\arraystretch}{0.5}
\caption{GPD parameters estimation, using the G-E-GPD algorithm , for the same example ($\boldsymbol{\theta}=[2,1,5,0.5]$, and $\beta=2.5$) and for $n=10^5$ : comparison with the ML and the PWM methods.} \label{tab2}
\centering
\scalebox{.9}{
\begin{tabular}{|c|c|c|c|c|}
\cline{3-5}
\multicolumn{2}{c|}{ } & G-E-GPD algorithm  & ML & PWM \\ \hline
\multirow{4}{*}{\vspace{0.75cm}$\xi=0.5$ }  & $\widetilde{\xi}$ &$0.5001$ & $0.5003$ & $0.5025$ \\  \cline{2-5}
\multirow{4}{*}{$\beta=2.5$ } & MSE$_{\xi}$ &  $1.1$ $10^{-5}$ & $1.73$ $10^{-4}$ & $5.08$ $10^{-4}$ \\  \hline
 & $\widetilde{\beta}$  &  $2.499$ & $2.5003$ & $2.4968$ \\  \cline{2-5}
    & MSE$_{\beta}$ & $4.08$ $10^{-4}$ & $1.37$ $10^{-3}$& $2.35$ $10^{-3}$ \\ \hline
\end{tabular}
}
\end{table}

\section{APPLICATION OF THE SELF-CALIBRATING METHOD ON REAL DATA}
\label{sec:Applic}


Once the performance of the algorithm  is validated on generated data, we apply it on real  data, considering two different domains: neuroscience and finance. Those data are essentially symmetric around the mean (that is why we keep the Gaussian component to model the mean behavior; nevertheless in the case of skewed data, as for insurance claims, it would be natural to replace the Gaussian component with a lognormal one, as already pointed out), and not necessarily independent. To underline the good performance of the self-calibrating method, for each application we compare, in terms of goodness-of-fit of extremes,  the results obtained with this method to those provided by standard EVT approaches: the graphical Mean Excess Plot (MEP) (see e.g. \cite{Embrechts1997}), and Hill (see \cite{Hill}) method (or the QQ-one  (see \cite{Kratz1996}), whenever the Hill plot seems somewhat inconclusive).

\subsection{Neuroscience: Neural Data}

\begin{figure}[H]
  \centering
  \includegraphics[scale=0.35]{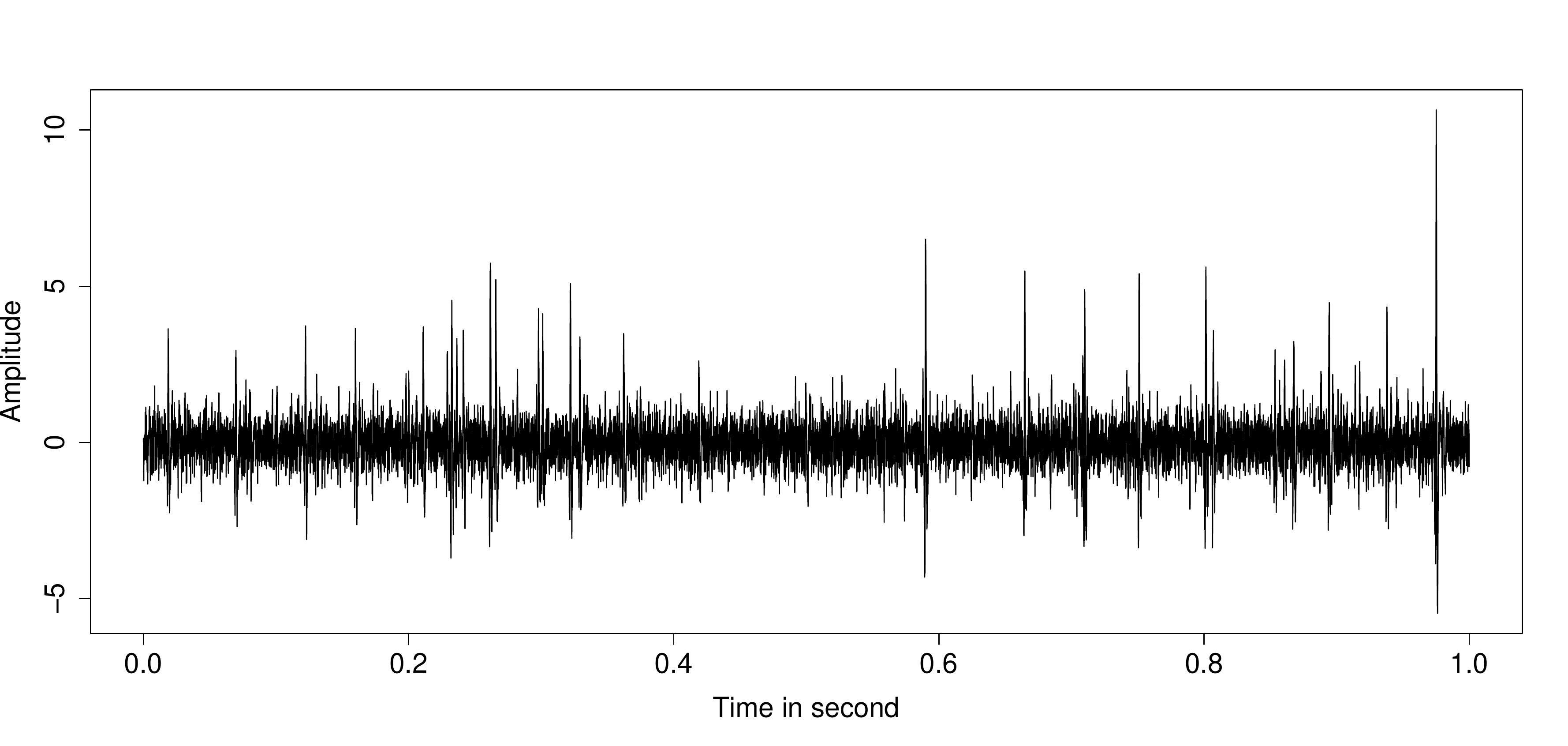}\\
  \caption{One second of neural data, extracellularly recorded.}\label{neural}
\end{figure}

Here we consider the data corresponding to twenty seconds, equivalent to $n=3\times10^5$ observations, of real extracellular recording of neurons activities, available in \cite{Pouzat_data} and measured (\emph{via} a microelectrode) on the antennal lobe of an adult locust.
The information to be extracted from these data, one second of which is represented  in Figure \ref{neural}, is the presence of action potentials \emph{viz} spikes (see e.g. \cite{MBOUP12}), which lies on the extreme behaviors (left and right) of the data. As shown in Figure \ref{neural},  the recorded data is corrupted by noise (mean behavior). This noise  corresponds mainly  to the activities of remote neurons w.r.t. the microelectorode (unuseful information). Hence the need to separate extremes (action potentials) from noise. The presence of extremes and the distinction between a correct detected spike and a false alarm (noise)  have been studied in \cite{Debbabi2012b}, when modelling only the right tail of a transformed neural data, by a GPD. Here we propose to complete the study, modelling the whole neural data and not only the extreme behavior.
\vspace{-4ex}
\vspace{-2ex}
Since the neural data can be considered  more or less as symmetric, we will evaluate the right side of the distribution with respect to its mode.
In what follows, we compare the results of (right side) neural data  fitting when using the self-calibrating, MEP, Hill and  QQ - methods, respectively.

\begin{itemize}
\item{\it Application of our self-calibrating method}

Applying the self-calibrating algorithm on the neural data set to model its right side, we obtain the following estimate of $\boldsymbol{\theta}$: $\widetilde{\boldsymbol{\theta}}=[-0.0681, 0.6297, 1.0301, 0.5398]$. In Figure~\ref{appli1} (1st row), we can see well, on a log-scale, the good fit of the  estimated hybrid cdf compared to the empirical one (see plot (a)), but also that of the right tail distribution (see plot (b)). We observe here that the exponential distribution is not needed for a good modelling of the data. Indeed, the two junction points overlap:  the estimates $\widetilde{u}_1$, $\widetilde{u}_2$ of $u_1$ and $u_2$, respectively, are very close to each other (with a distance equal to $ 4.3268$ $10^{-5}$) (see plot (a)).  It confirms what has already been observed in \cite{Debbabi2015b} using a two-component model.

\begin{figure}[htp]
\centering
\subfigure [Right side of the empirical cdf versus the hybrid cdf obtained with the self-calibrating method]{
\includegraphics [ height =4.25cm,width=7cm]{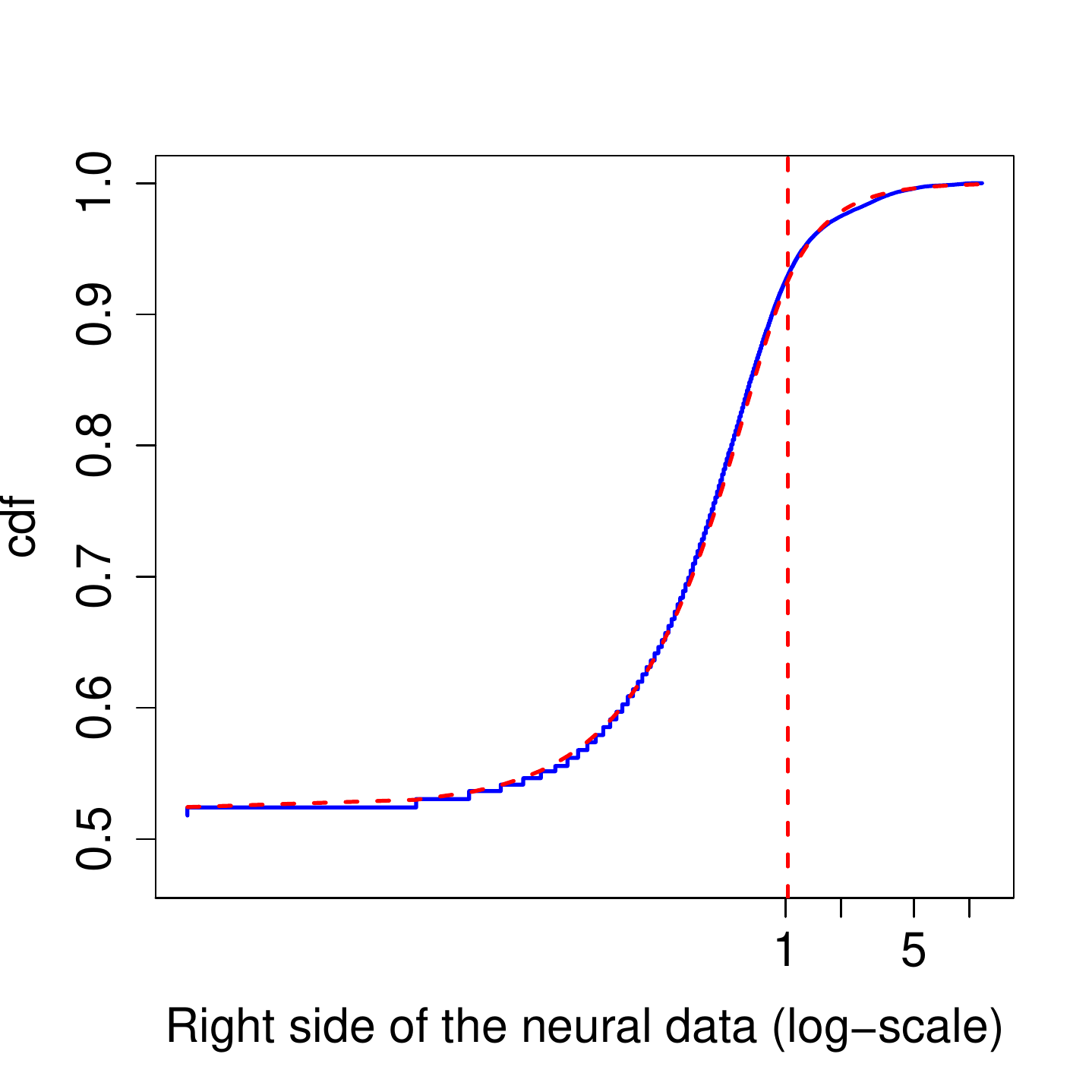}
}
\quad
\subfigure [Empirical tail distribution versus GPD. \hspace{2cm} Self calibrating method]{
\includegraphics [ height =4.25cm,width=7cm]{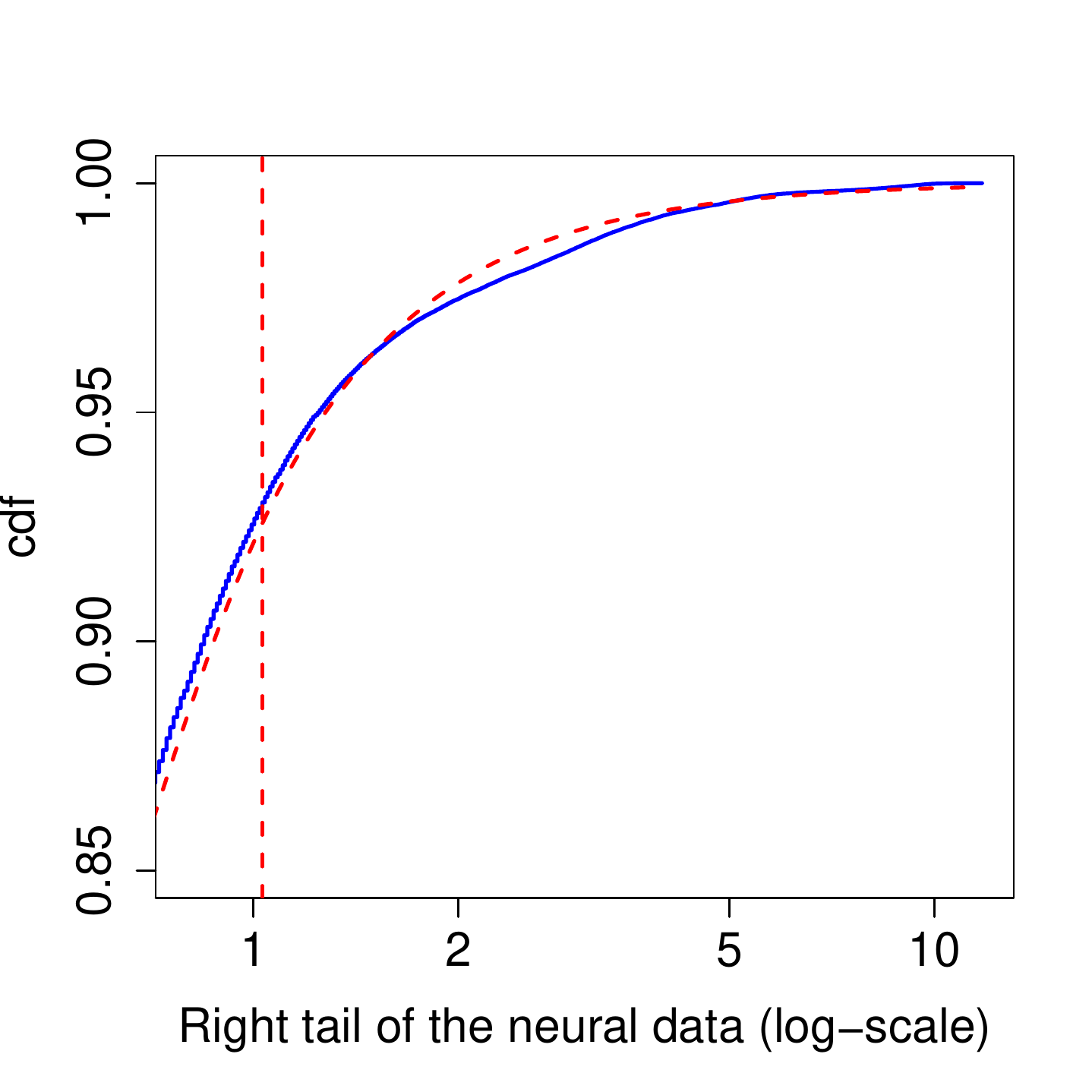}
}
\\
\quad
\subfigure [Mean Excess Plot]{
\includegraphics [ height =4.25cm,width=7cm]{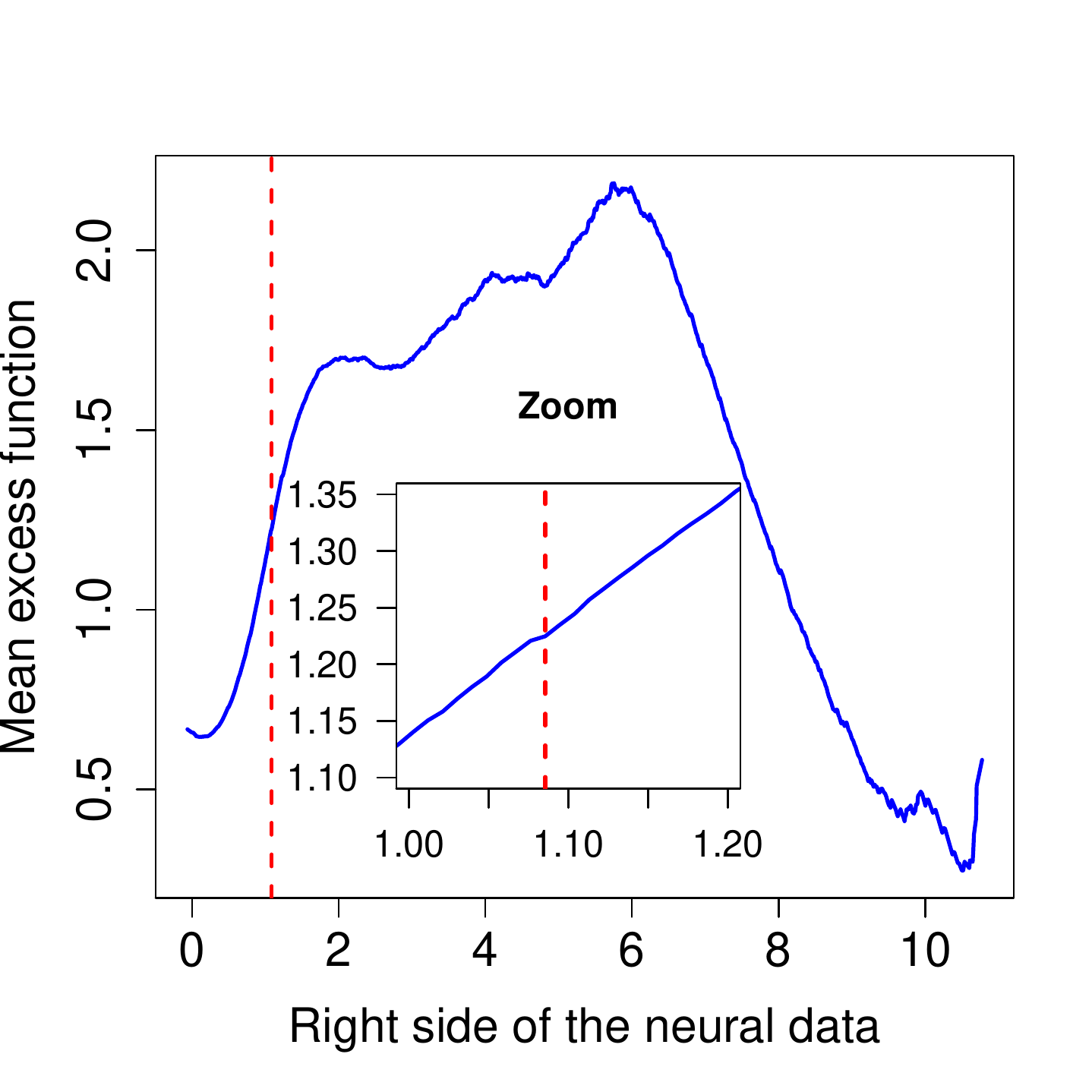}
}
\quad
\subfigure [Empirical tail distribution versus GPD. \hspace{2cm} MEP \& PWM methods]{
\includegraphics [ height =4.25cm,width=7cm]{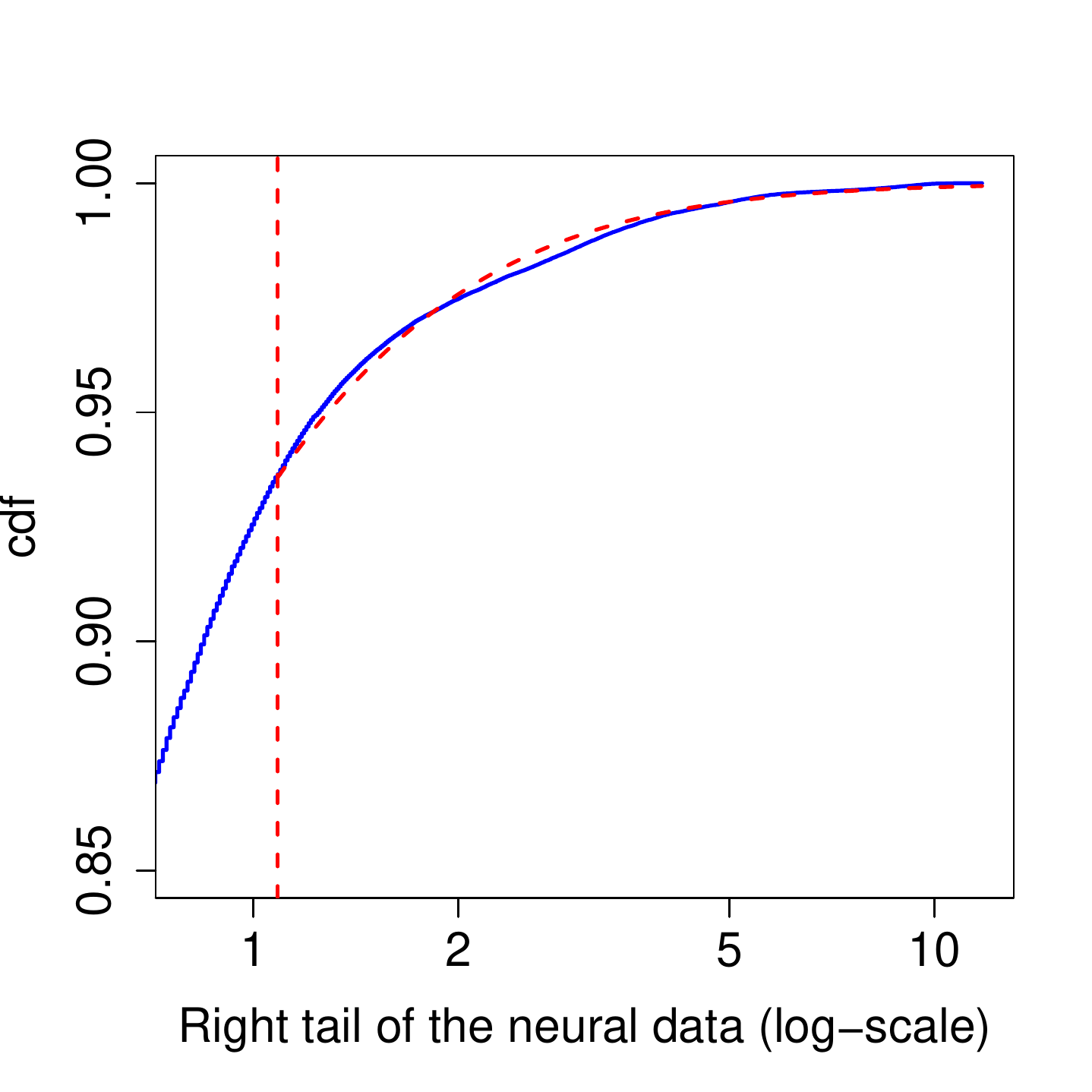}
}
\\
\quad
\subfigure [Hill plot]{
\includegraphics [ height =4.25cm,width=7cm]{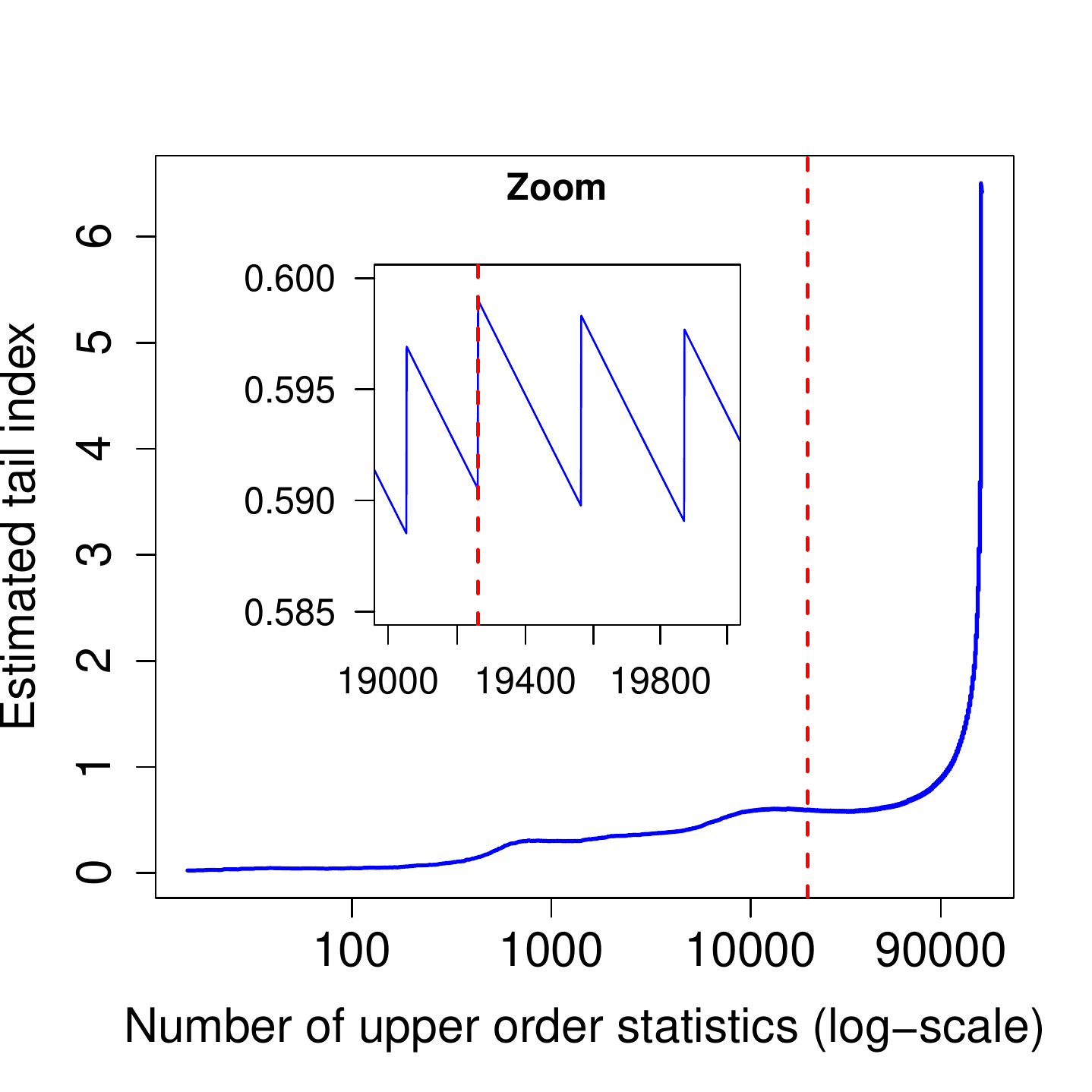}
}
\quad
\subfigure [Empirical tail distribution versus GPD. \hspace{2cm} Hill method]{
\includegraphics [ height =4.25cm,width=7cm]{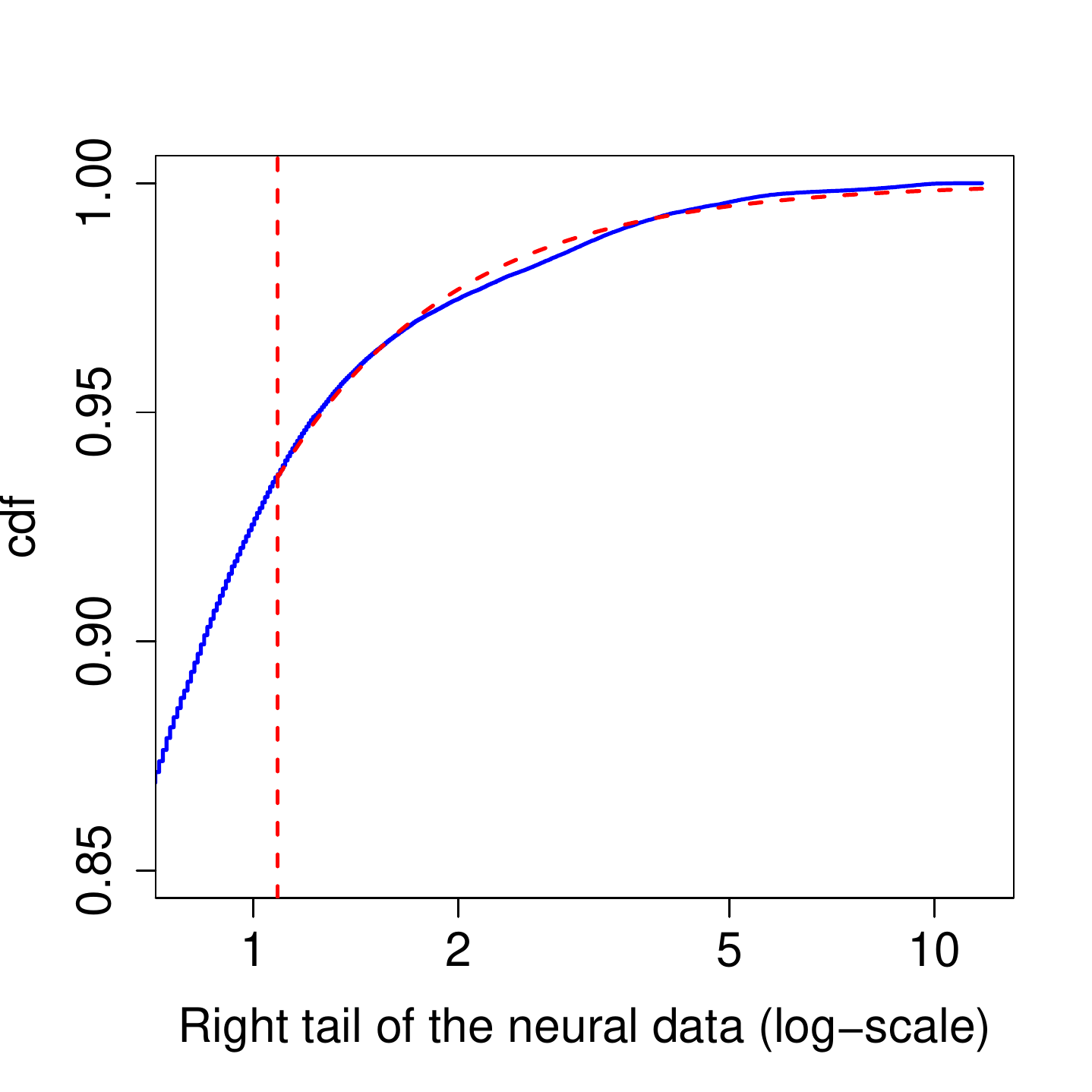}
}
\\
\quad
\subfigure [QQ-estimator plot]{
\includegraphics [ height =4.25cm,width=7cm]{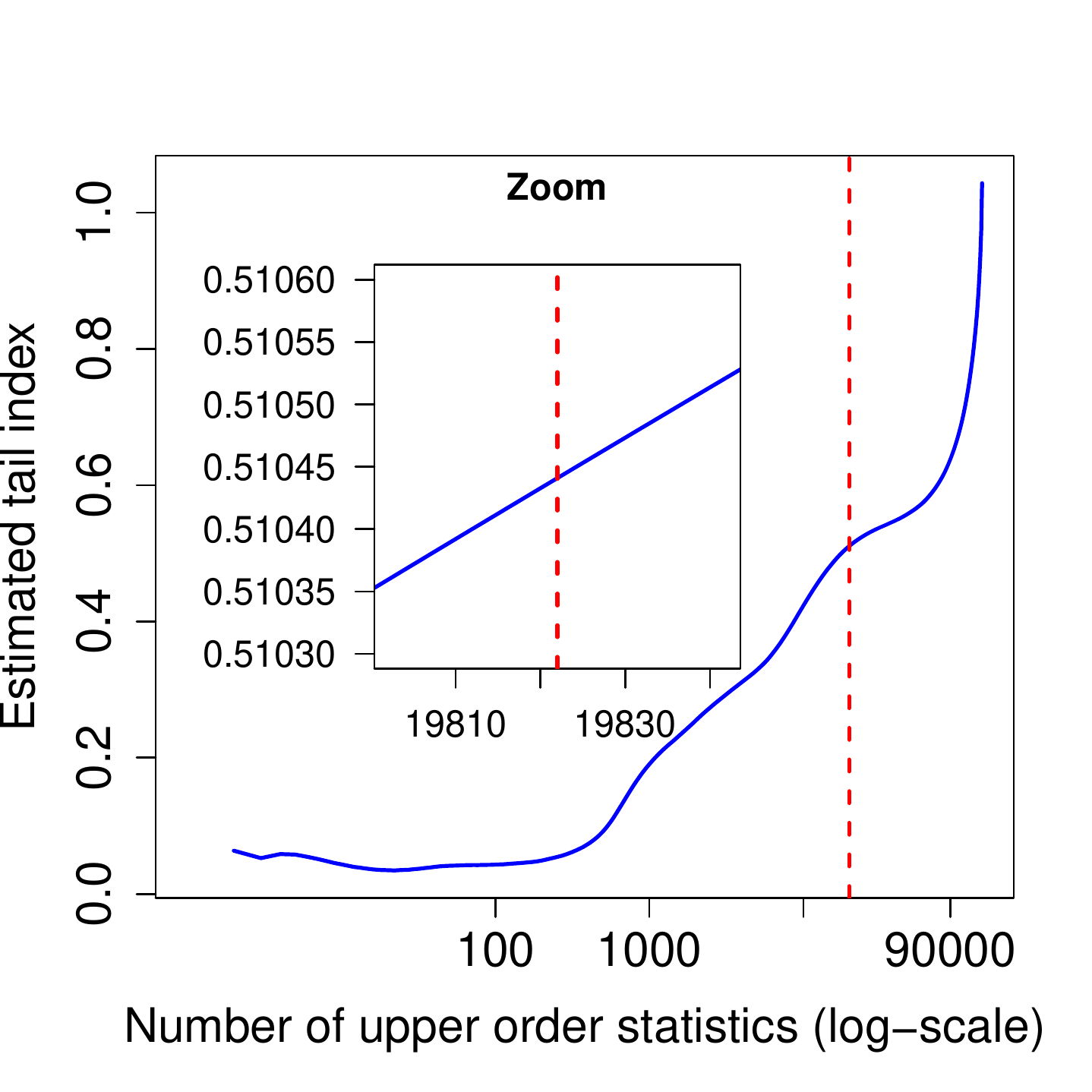}
}
\quad
\subfigure [Empirical tail distribution versus GPD. \hspace{2cm} QQ method]{
\includegraphics [ height =4.25cm,width=7cm]{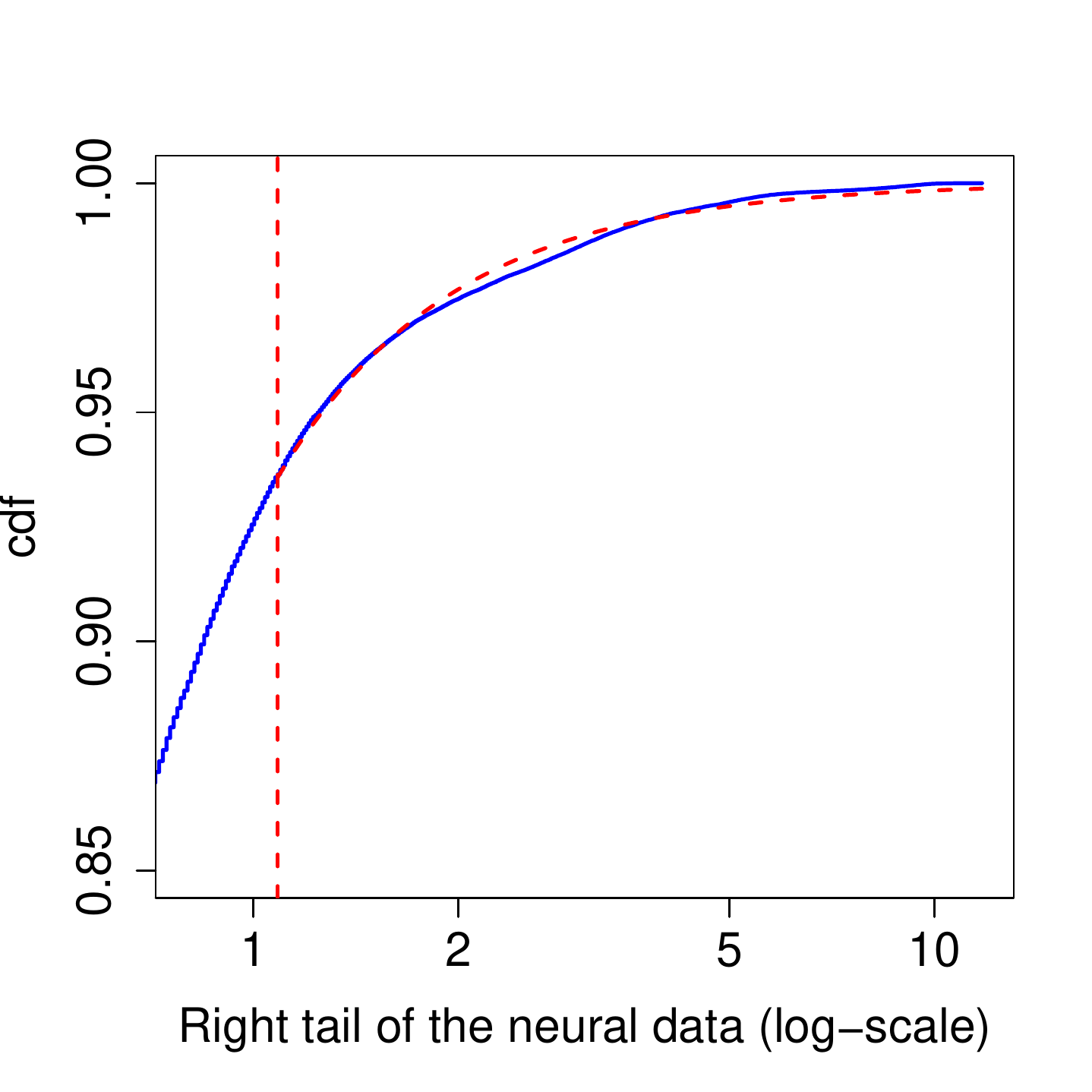}
}
\caption{Extremes modelling using different methods. For each plot, the (blue) continuous  curve is empirical (even for zoomed curves), while the (red) dashed curve and  vertical line represent the estimated GPD and threshold, respectively, using the  associated method.}
\label{appli1}
\end{figure}

\item {\it Application of the MEP method}

We draw the Mean Excess Plot (MEP) (see plot (c)) to manually determine the threshold above which data are Generalized Pareto distributed (that we refer as the GPD threshold). We look  from which threshold (high enough to match the theory, but not too high to have enough observations) the MEP behaves linearly. Then, we estimate  the  corresponding GPD  parameters using, for instance, the  PWM method. Several values of the threshold have been selected. We choose the one offering the smallest MSE between the empirical tail distribution and the estimated GPD (see the zoomed part of plot (c), where the linear behavior of the MEP is pointed out). The reliability of this graphical method, in terms of goodness-of-fit of extremes above the selected threshold, is illustrated  in plot (d).

\item {\it Application of the Hill method or/and the QQ-method}

In a similar way, we determine the GPD threshold graphically  from the Hill plot (see plot (e)), representing the Hill estimator of the GPD tail index as a function of  the number of the upper order statistics. Because of its volatile behavior (as observed  in the corresponding zoomed plot), we also provide the plot for the QQ-estimator (see plot (g)) to confirm the threshold detection.
After several tests, we select the number of upper order statistics above which we observe a stability of the Hill plot (see plot (e)), or a linear behavior for the plot of the QQ-estimator of the tail index (see plot (g)). The associated threshold minimizes the MSE between the estimated and the empirical tail cdf. We note that once the number of upper order statistics is selected, the associated threshold and tail index are determined, and the scale parameter is estimated as the product of this threshold by the tail index. We draw in plot (f), (h) respectively, the empirical tail distribution and the estimated GPD.

\end{itemize}

{\it Comparison of the results obtained via the various methods}

\begin{figure}[H]
  \centering
  \includegraphics[scale=0.35]{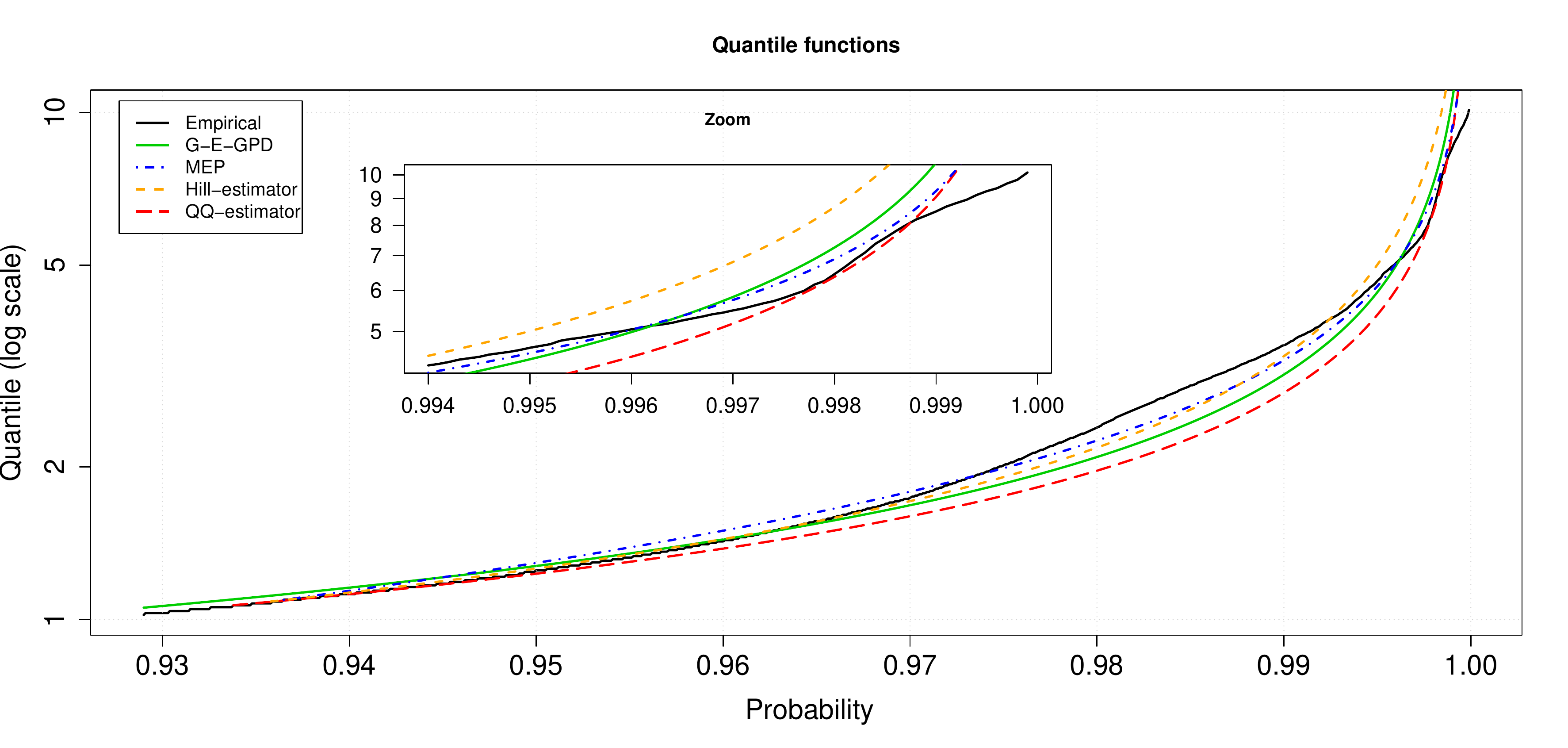}
  \caption{Neural data: Comparison between the empirical quantile function and the estimated ones via the self-calibrating method and the graphical methods.}\label{qneural}
\end{figure}

In Table \ref{tab3}, we present the results obtained with the self-calibrating method, the MEP, Hill and  QQ methods. Since the three graphical approaches fit only the tail distribution, the comparison of the methods will focus on the goodness-of-fit of the GPD component. 
%
 As observed in this table, the MSE between the estimated cdf and the empirical one, using only data above the selected threshold, is small enough for the four methods ensuring a reliable modelling of extremes. The GPD threshold and the estimated tail index are of the same order of magnitude for all methods; it confirms  that our algorithm works in the right direction.
 We can also notice the good performance of these methods through Figure \ref{qneural}, where we plot  the empirical  quantile function and the estimated ones using the self-calibrating method and the various graphical ones.  However, the advantage of our method is that it is unsupervised, {\it i.e.} it does not need the intervention of the user to select the threshold manually. Moreover it provides a good fit  between the hybrid cdf estimated on the entire data sample (the right side for this data set) and the empirical cdf, with a MSE of order $10^{-5}$.
 \begin{table}[H]
\renewcommand{\arraystretch}{.5}
\caption{Comparison between the self-calibrating method and the three graphical methods: MEP, Hill and QQ ones. $N_{u_2}$ represents the number of observations above $u_2$. The distance gives the MSE between the empirical (tail or full respectively) distribution and the estimated one from a given model (GPD or hybrid G-E-GPD respectively). The neural data sample size is $n=3\times10^5$.}\label{tab3}
\centering
\scalebox{.9}{
\begin{tabular}{|c|c|c|c|c|c|}
  \hline
Model & tail index  &  threshold  & $N_{u_2}$ & distance  & distance \\
      & ($\xi$) &  ($u_2$) &  & (tail distr.) &  (full distr.)\\
  \hline
   GPD  & MEP (PWM): $0.3326$ &  $1.0855=q_{_{93.64\%}}$ &19260& $3.26$ $10^{-6}$ &\\
   \hline
     GPD  & Hill-estimator: $0.599$ &  $ 1.0855=q_{_{93.64\%}}$ &19260& $2.07$ $10^{-6}$ &\\
   \hline
 GPD  & QQ-estimator: $0.5104$ &  $ 1.0671=q_{_{93.47\%}}$ &$19871$& $1.26$ $10^{-5}$ &\\
   \hline
 G-E-GPD & Self-calibrating method: $0.5398$& $1.0301=q_{_{92.9\%}}$& 21272 & $7.79$ $10^{-6}$ & $9.31$ $10^{-5}$\\
   \hline\hline
  \end{tabular}
  }
  \end{table}
  \vspace{-.75cm}
\subsection{Finance: S\&P500 log-returns}

The second application considered in this work concerns the S\&P500 
log-returns from January 2, 1987 to February 29, 2016, available in the tseries package (see \cite{tseries2016}) of the R programming language (see \cite{R2014}).  It is well known that
log-returns of financial stock indices exhibit left and right heavy tails, with a slight different tail index from one to the other. It is important in such context to evaluate the nature of tail(s) in order to compute the capital needed by a financial institution to cover their risk, often expressed as a Value-at-Risk ({\it i.e.} a quantile) of high order.  We check on these S\&P500 data how our method performs, comparing its evaluation of the tail indices with e.g. the standard Hill estimates. It also delivers, thanks to the full distribution fit, a way to compute the expected equity premium (/return) ) with respect to the interest rates. This is important e.g. when optimizing a portfolio composed of both bonds and equity.

\begin{figure}[H]
  \centering
  \includegraphics[scale=0.35]{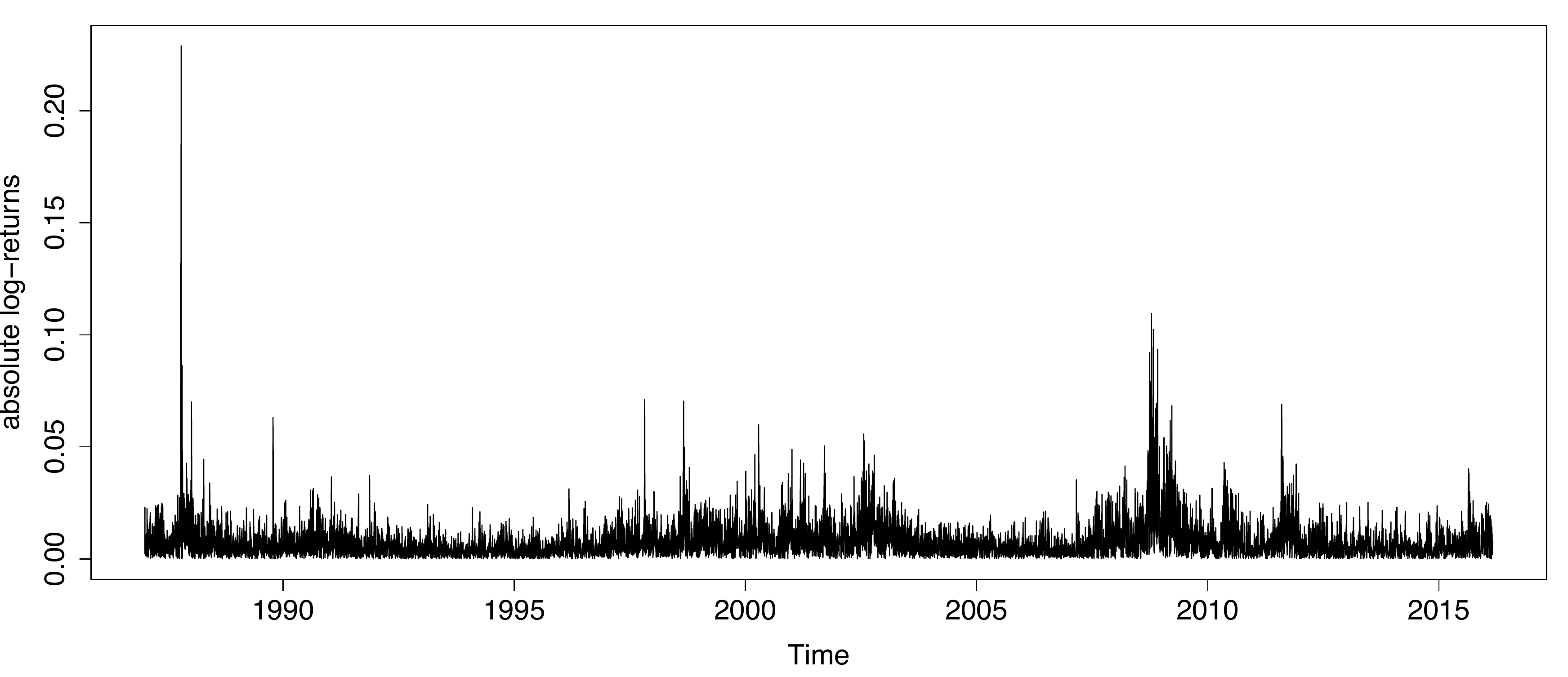}
  \caption{S\&P500 absolute daily log-returns from January 2, 1987 to February 29, 2016.}\label{sp500}
\end{figure}

First let us look at the absolute value of the S\&P500 log-returns, corresponding to $n=7348$ observations represented in Figure \ref{sp500}.
 It is well known that the absolute value of financial returns are autocorrelated, but also that their extremes are not (for a thorough discussion of this point and empirical evidences, see \cite{Hauksson2001}). In time of crisis, as e.g. in 2008-09, we observe an increase of the dependence between various financial indices, in particular in the extremes. This is to be distinguished from a dependence of the extremes within a univariate financial index, which is not observed (\cite {Hauksson2001}).

As for the neural data, we apply our self-calibrating method and the graphical ones (MEP and QQ) for comparison.  Note that we display only the plots associated with the QQ-method and not the Hill one, since the QQ-threshold is easier to detect, as already commented. Nevertheless, we provide the numerical results for both Hill and QQ methods, using the $\sqrt{n}$ upper order statistics to compute the Hill estimator (this selected threshold has been empirically shown to be relevant for financial data in \cite{Blum2003}).

\begin{figure}[htp]
\centering
\subfigure [Empirical cdf versus the hybrid cdf obtained with the self-calibrating method]{
\includegraphics [ height =5cm,width=7cm]{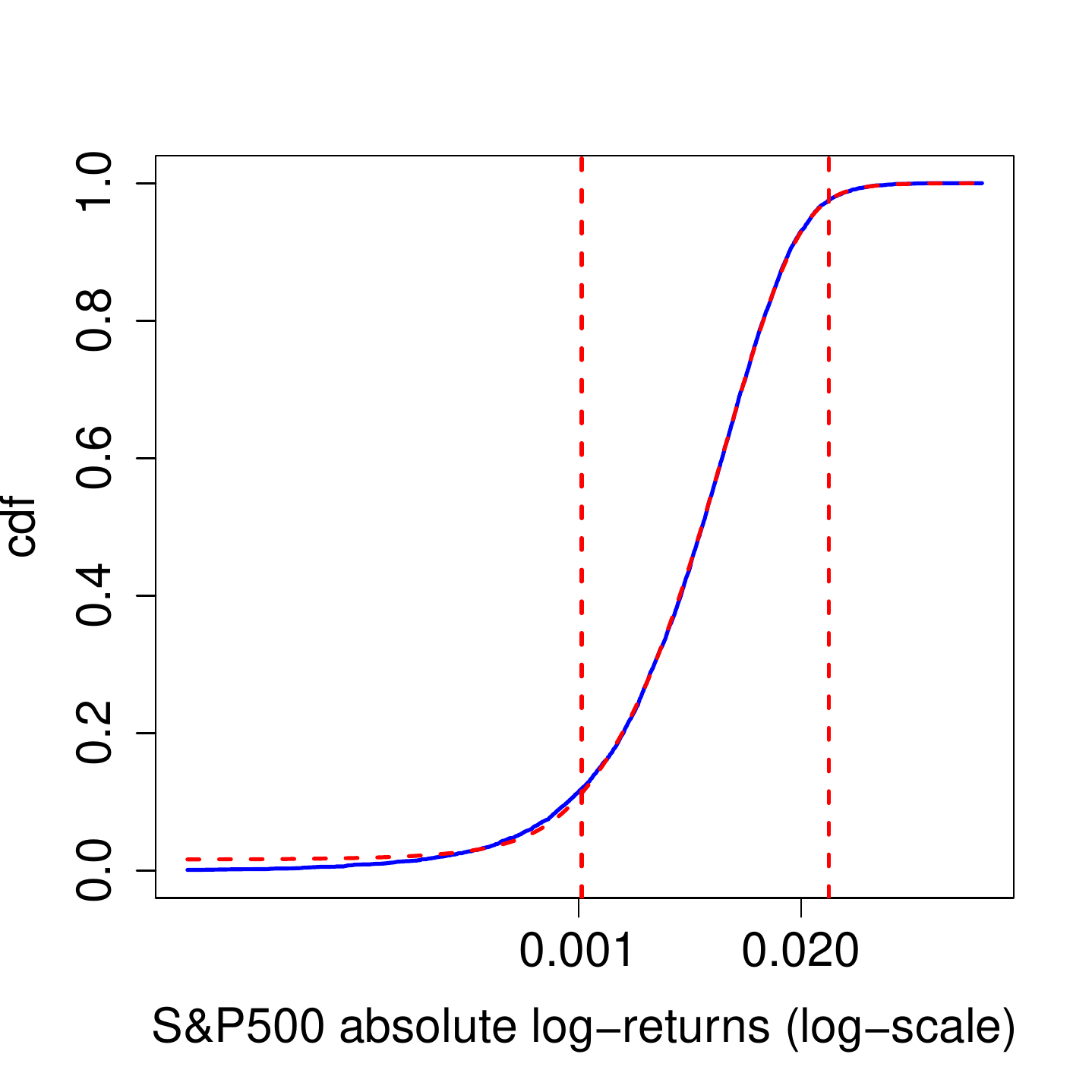}
}
\quad
\subfigure [Empirical tail distribution versus GPD. \hspace{2cm} Self calibrating method]{
\includegraphics [ height =5cm,width=7cm]{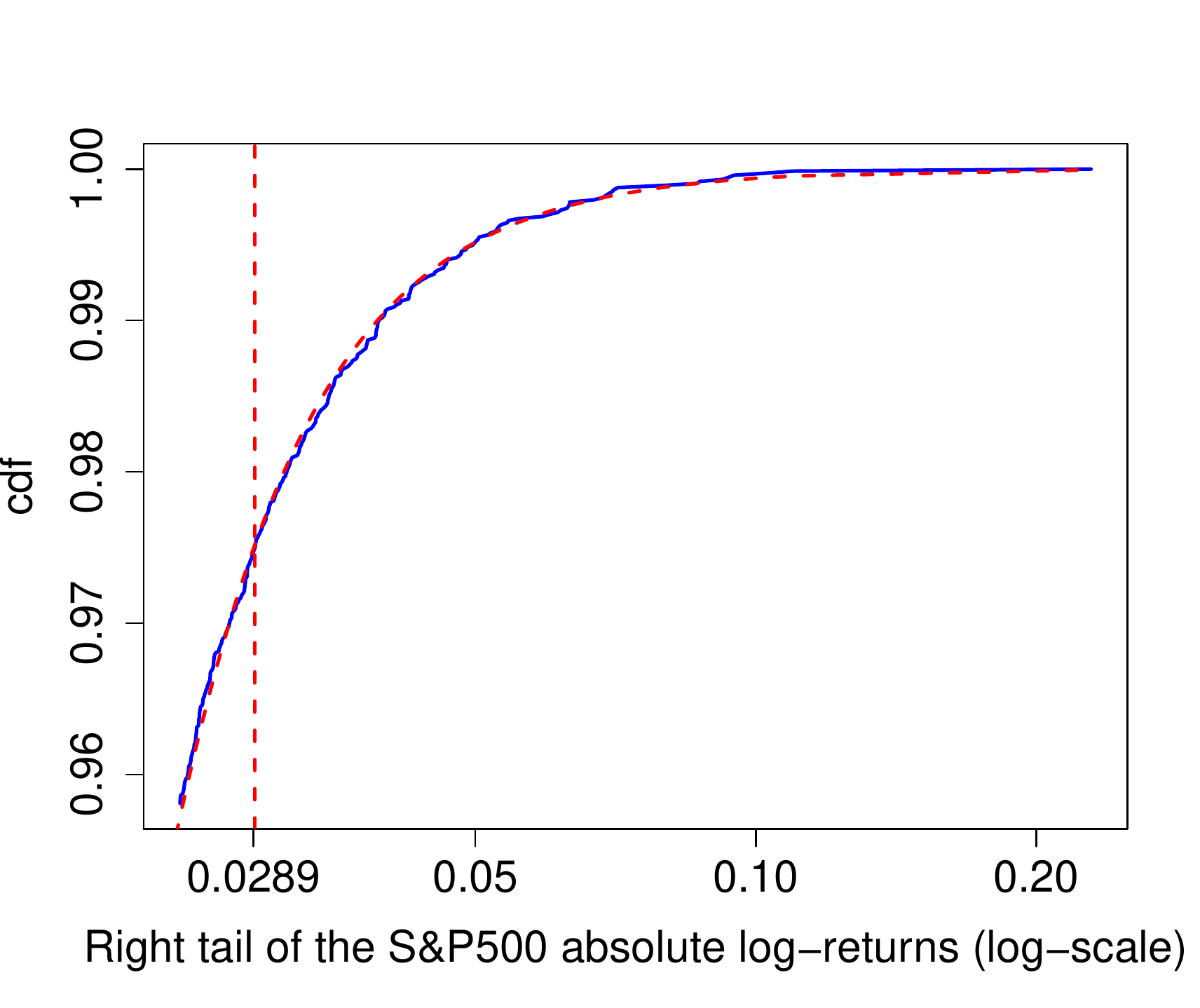}
}
\vspace{-3ex}
\subfigure [Mean Excess Plot]{
\includegraphics [ height =5cm,width=7cm]{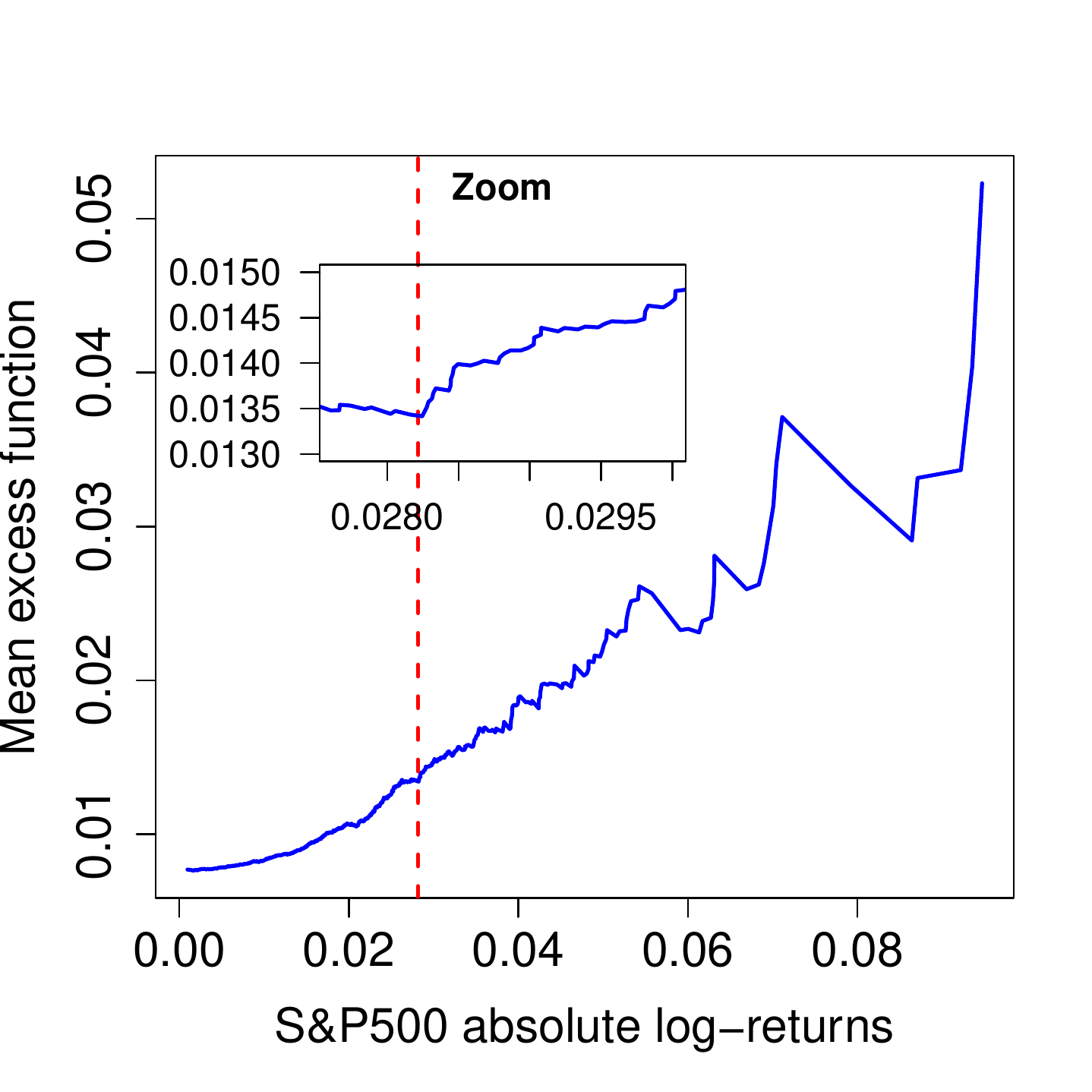}
}
\quad
\subfigure [Empirical tail distribution versus GPD. \hspace{2cm}  MEP \& PWM methods]{
\includegraphics [ height =5cm,width=7cm]{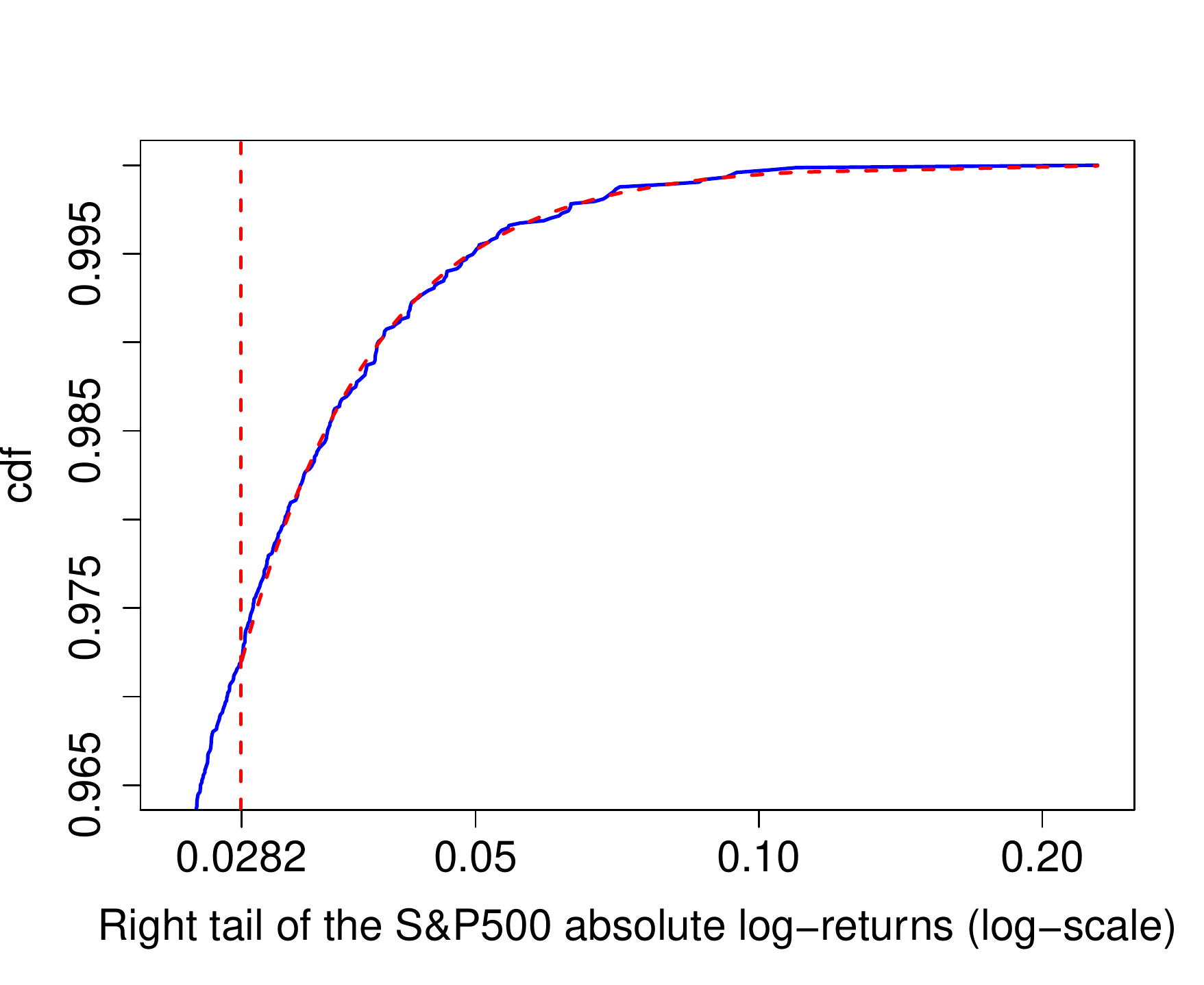}
}
\vspace{-1ex}
\subfigure [QQ-estimator plot]{
\includegraphics [ height =5cm,width=7cm]{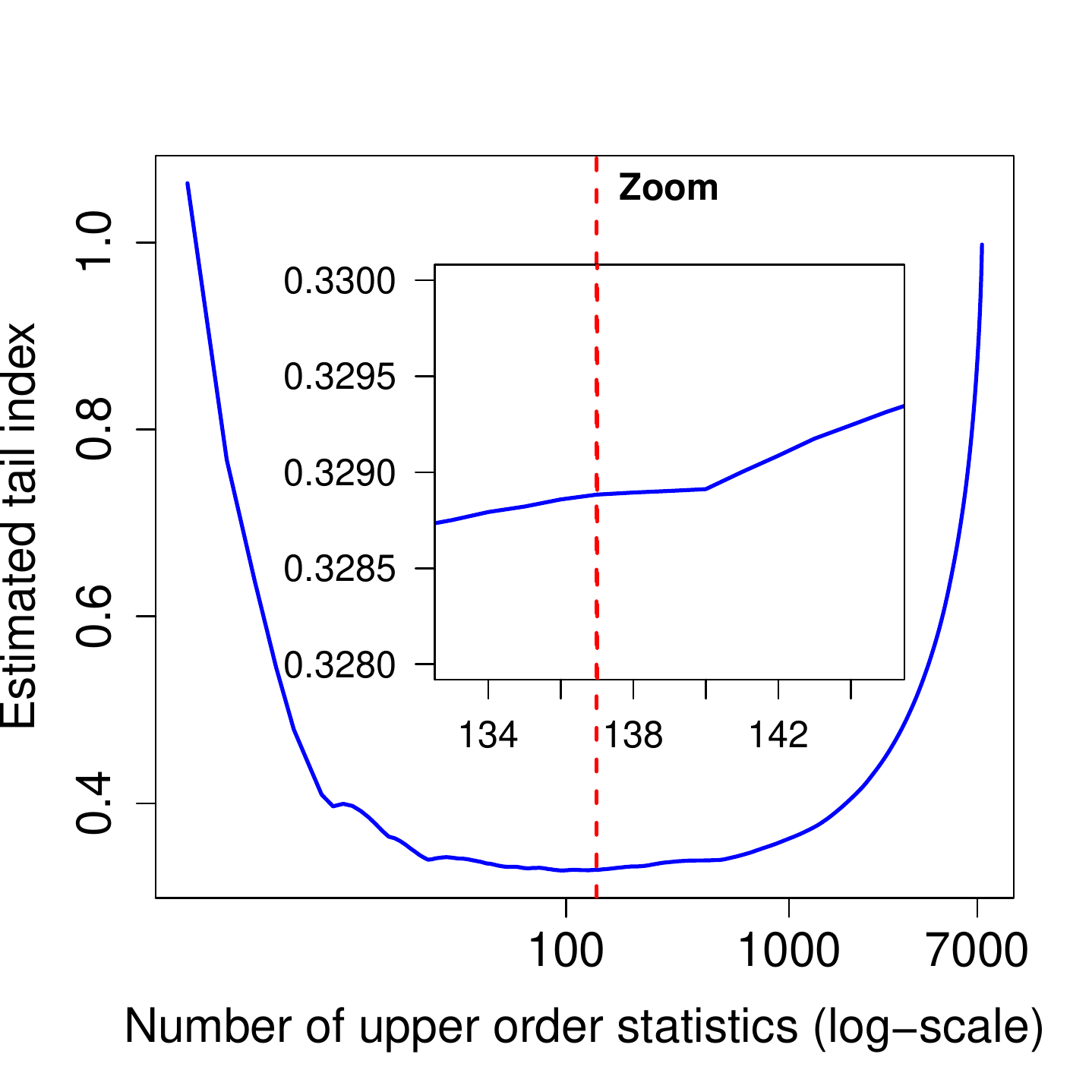}
}
\quad
\subfigure [Empirical tail distribution versus GPD. \hspace{2cm} QQ method]{
\includegraphics [ height =5cm,width=7cm]{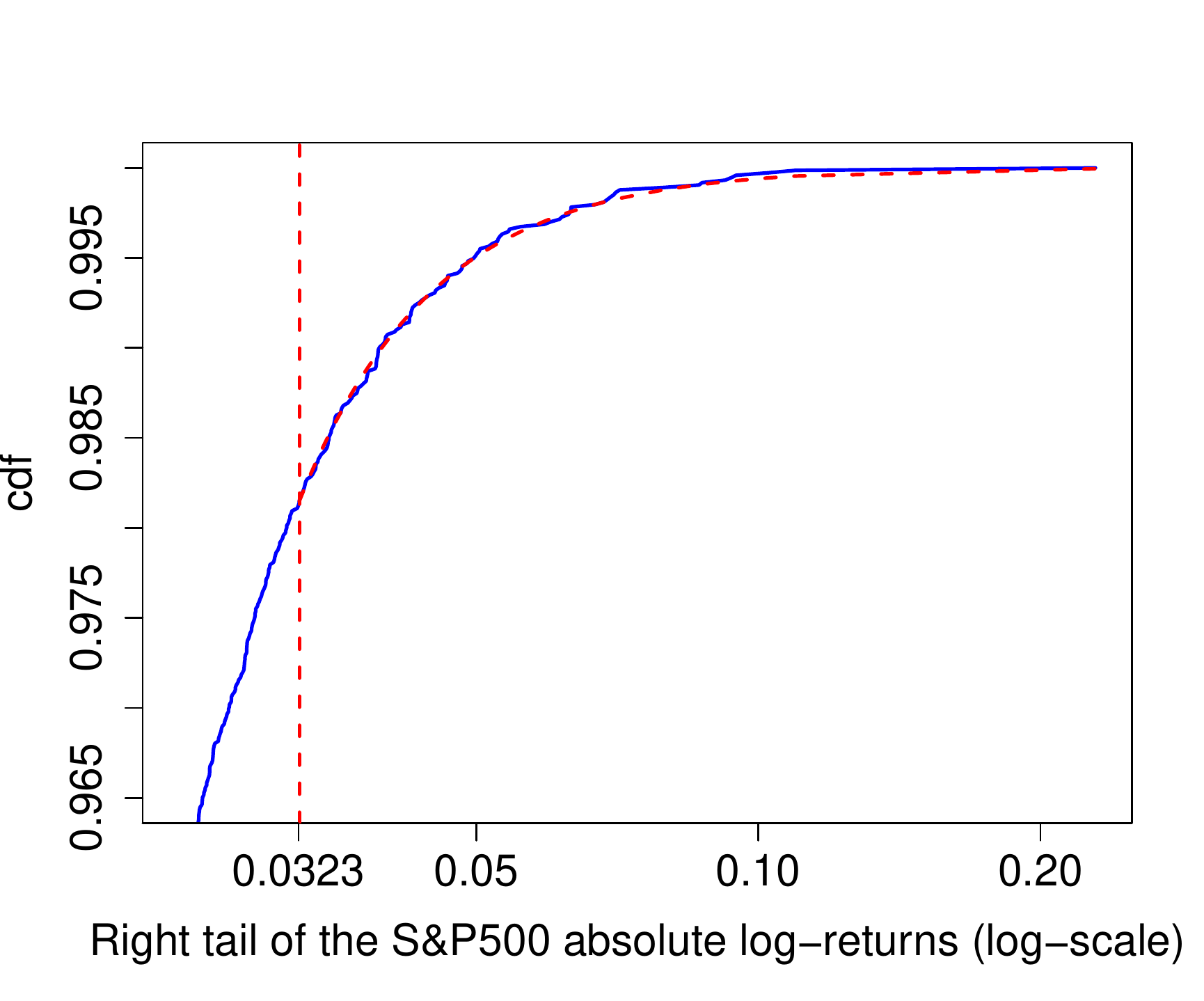}
}
\vspace{-1ex}
\caption{Extremes modelling using different methods. For each plot, the (blue) continuous  curve is empirical (even for zoomed curves), while the (red) dashed curve and  vertical line represent the estimated GPD and threshold, respectively, using the  associated method.}
\label{appli2}
\end{figure}

 The results are illustrated  in Figure \ref{appli2}. In plot (a), we draw (on a log-scale) the empirical cdf and the hybrid one obtained via our self-calibrating method, where the two vertical dashed lines represent the two junction points of the hybrid model. The related right tail fit is given in plot (b).

The  MEP and the selected threshold are given in plot (c) (the zoomed part shows the MEP  linear behavior above the selected threshold), while the corresponding extremes fit is given in plot (d), the GPD parameters being estimated by the PWM method.

 Finally, the QQ-estimator plot and the selected number of upper order statistics  are represented in plot (e), with a zoom illustrating the linear behavior of the QQ-estimator plot above the selected number of upper order statistics. The  corresponding extremes fit is shown in plot (f).

\begin{figure}[H]
  \centering
  \includegraphics[bb= 0 17 995 462,scale=0.35]{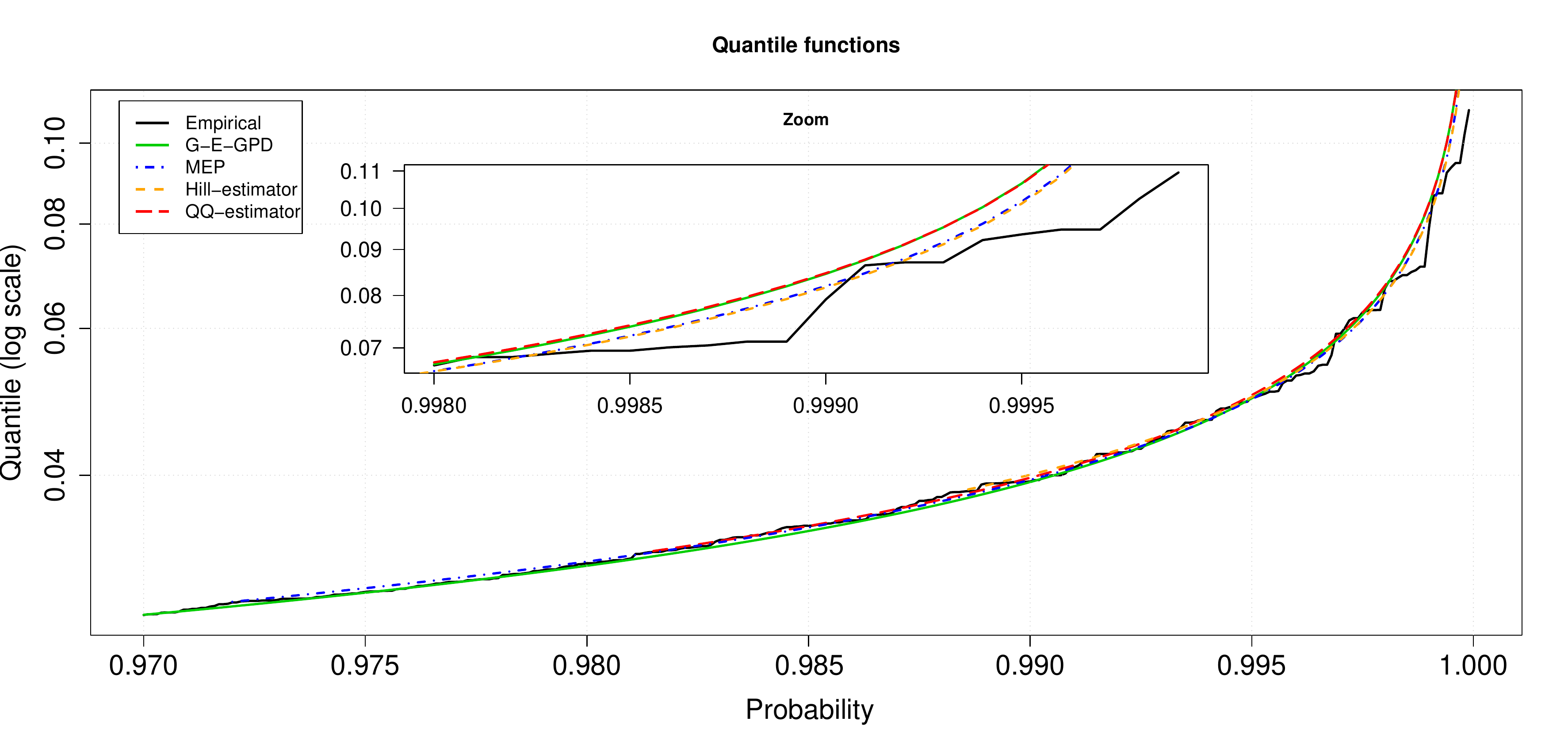}\\
  \caption{S\&P 500 absolute log-returns data: Comparison between the empirical quantile function and the estimated ones via the self-calibrating method and the graphical methods.}\label{qSP500}
\end{figure}

 The numerical results obtained for the threshold and tail index, as well as for the MSE between the empirical tail distribution and the estimated GPD using the four methods respectively, are reported in Table \ref{tab4}.  We can notice that all methods offer a good fit of the tail distribution, with a slightly overestimation for the G-E-GDP and QQ methods compared with the MEP and Hill ones.
\begin{table}[H]
\renewcommand{\arraystretch}{.5}
\caption{Comparison between the self-calibrating method and the three graphical methods: MEP, Hill and QQ ones.  The S\&P 500 absolute log-returns data sample size is $n=7348$.} \label{tab4}
\centering
\scalebox{0.9}{
\begin{tabular}{|c|c|c|c|c|c|}
  \hline
Model & tail index  &  threshold  & $N_u$ & distance  & distance \\
      & ($\xi$) &  ($u_2$) &  & (tail distr.) &  (full distr.)\\
  \hline
  GPD  & MEP: $0.3025$ &  $0.0282=q_{_{97.21\%}}$ &206& $1.78$ $10^{-7}$ &\\
   \hline
   GPD  & Hill-estimator: $0.3094$ &  $ 0.0382=q_{_{98.85\%}}$ &85& $ 4.49$ $10^{-8}$ &\\
   \hline
 GPD  & QQ-estimator: $0.3288$ &  $0.0323=q_{_{98.14\%}}$ &137& $6.01$ $10^{-8}$ &\\
   \hline
 G-E-GPD & Self-calibrating method: $0.3332$ & $0.0289=q_{_{97.49\%}}$& 184 & $ 1.95$ $10^{-7}$ & $1.06$ $10^{-5}$\\
   \hline
  \end{tabular}
  }
  \end{table}
In Figure \ref{qSP500}, we also give a comparison of the estimated quantile function using the G-E-GPD method and the graphical (MEP, Hill and QQ) ones.

In Table \ref{tab4} and Figure \ref{qSP500}, we observe once again similar results for the various methods. It confirms the good performance of the self-calibrating method to estimate the tail distribution. As already said, this latter method also provides a good modelling for  the entire cdf.

\begin{figure}[H]
  \centering
  \includegraphics[scale=0.35]{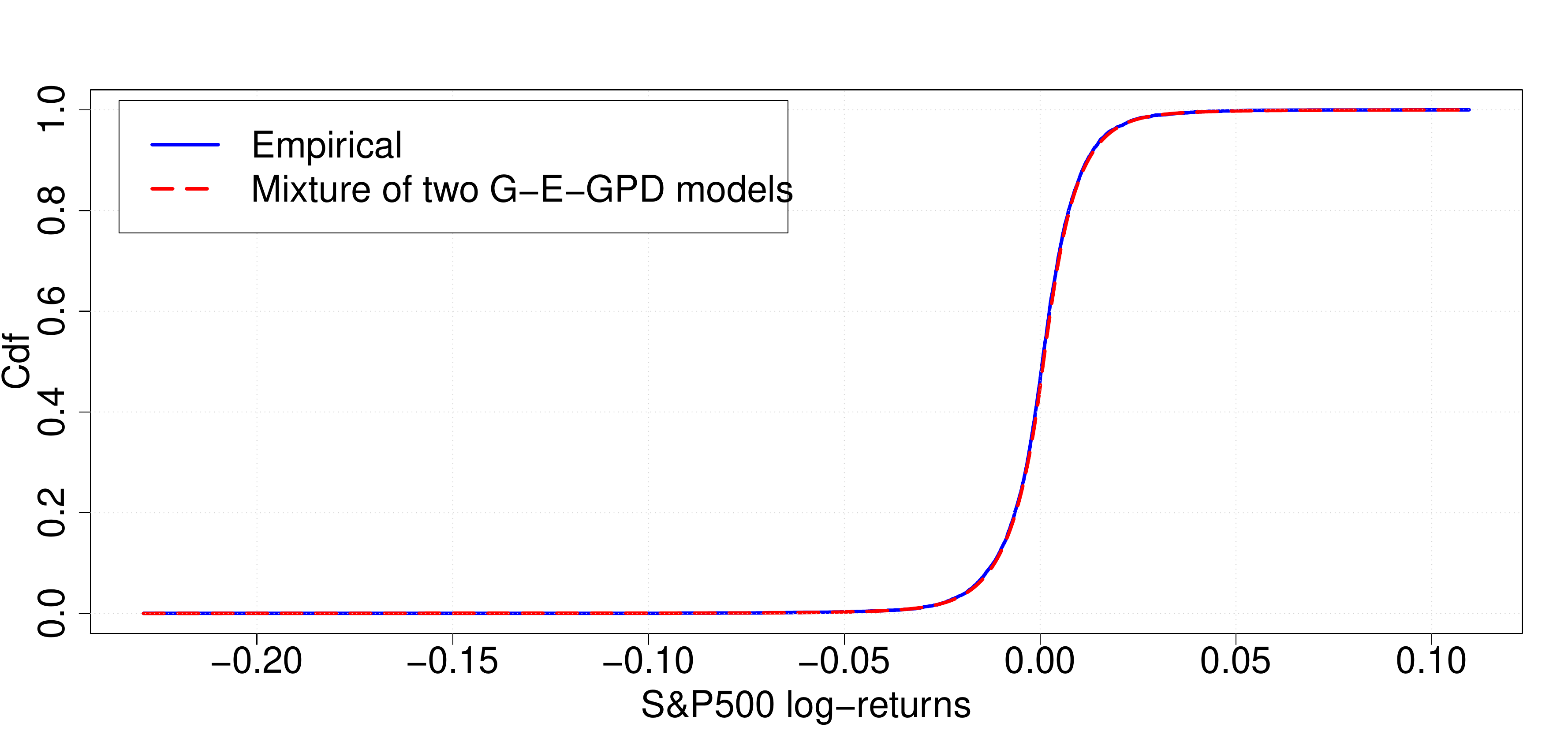}
  \caption{The cdf fit of the S\&P500 log-returns data  using a mixture of two G-E-GPD hybrid models.}\label{sp500cdf}
\end{figure}

Now, for completeness, we apply the self-calibrating algorithm to model the upper and lower tails, separately, of the S\&P 500 log-returns (say $X$).
 For the upper tail, we consider the vector of parameters  $\boldsymbol{\theta_2}$, minimizing the MSE between empirical and theoretical cdfs for only positive observations. In the same way, we determine the vector of parameters $\boldsymbol{\theta_1}$, related to the lower tail (considering -X, when applying the algorithm). The results are given in Table \ref{tab5}, where we compare them with e.g. the  Hill estimators (using the $\sqrt{n}$ upper order statistics). As expected, similar results are observed.
\begin{table}[H]
\renewcommand{\arraystretch}{.5}
\caption{ Comparison between the self-calibrating method and the Hill  one, applied to S\&P 500 log-returns data for both tails (lower and upper).} \label{tab5}
\centering
\scalebox{0.9}{
\begin{tabular}{|c|c|c|c|c|}
  \hline
Model &  tail index  &  threshold  & $N_u$ & distance \\
      & ($\xi$) &  ($u_2$) &  & (tail distr.) \\
       \hline
\multicolumn{5}{c}{ Lower tail}\\
 \hline
   GPD  & Hill-estimator: $0.3597$ &  $ 0.0301=q_{_{98.84\%}}$ &86& $6.4576$ $10^{-8}$ \\
   \hline
 G-E-GPD & Self-calibrating method: $0.3545$ & $0.02896=q_{_{98.63\%}}$& 100 & $2.6429$ $10^{-7}$ \\
   \hline
   \multicolumn{5}{c}{ Upper tail}\\
 \hline
   GPD  & Hill-estimator: $0.3225$ &  $ 0.0288=q_{_{98.84\%}}$ &86& $ 4.4285$ $10^{-7}$ \\
   \hline
 G-E-GPD & Self-calibrating method: $0.3360$ & $0.0266=q_{_{98.51\%}}$& 109 & $ 3.8955$ $10^{-7}$ \\
   \hline
  \end{tabular}
  }
  \end{table}
 We can then gather those results into a multi-component distribution to model the whole data reliably,  as illustrated in Figure \ref{sp500cdf}. This distribution corresponds to a mixture of the two G-E-GPD hybrid models used to estimate the upper tail and the lower one, with $\boldsymbol{\theta_1}$ and $\boldsymbol{\theta_2}$ as vector of parameters, respectively. The two hybrid distributions are linked together at $\mu=0$ (mean of the Gaussian components), where the continuity is imposed (note that in the general case, the junction point of the two hybrid models may be chosen as the data mode), and are weighted by $\alpha_1$ and $\alpha_2$, respectively, to obtain a distribution. This multi-component distribution can be expressed as $\displaystyle
h_{mix}(x)=\alpha_1 h(-x;\boldsymbol{\theta_1})\mathbbm{1}_{]-\infty,0[}(x)+\alpha_2h(x;\boldsymbol{\theta_2})\mathbbm{1}_{[0,+\infty[}(x)$,
$h$ representing the G-E-GPD hybrid distributions.

\section*{CONCLUSION}

In this paper, we propose a self-calibrating method to model heavy tailed data that may be non-homogeneous and multi-component. We develop it introducing  a general non-degenerate hybrid $\mathcal{C}^1$  distribution for heavy tailed data modelling, which links a normal distribution to a GPD via an exponential distribution that bridges the gap between mean and asymptotic behaviors.
The three distributions are connected to each other at junction points estimated by an iterative algorithm, as are the other parameters of the model. The convergence of the algorithm is studied analytically for one part and numerically for the other. The performance of the method is studied on simulated data. Based on the simulation results, we observe that our unsupervised algorithm offers a judicious fit of the 
right heavy tailed data, in particular with an accurate estimation of the tail index of the GPD that fits the extremes over the tail threshold.
Several applications of the method have been successfully performed on real data (in particular on insurance data, with practitioners). We give two of them on data coming from very different fields,  neural data and financial ones, accompanied by a comparison with other existing EVT methods.

Note that this method has been developed when considering 
right heavy tailed data; it can of course be applied in the same way when having  a left heavy tail, or when having a heavy tail on each side (without requiring a symmetry).

This method has many advantages and should become of great use in practice.
The main advantage is to be self-calibrating, avoiding the somehow arbitrary resort, when fitting the tail, to standard graphical methods (e.g. MEP or tail index estimation methods) in EVT.
A second advantage is to fit with the same iterative algorithm the full distribution of observed heavy tailed data, of any type whenever smooth enough ($\mathcal{C}^1$-distribution), providing an accurate estimation of the parameters for the mean and extreme behaviors.  It certainly answers a big concern encountered by  practitioners (as for instance in insurance, when pricing premiums (see e.g.   \cite{busse2014}, \S 2.1) for which both expectation and risk factor (measured in the tail of the distribution) are needed).
Moreover the method is quite general: besides the GPD needed when fitting the heavy tail, the other components might be chosen differently, not using limit behavior (CLT) but distributions chosen specifically for the data that are worked out (as e.g. lognormal for insurance claims). It would not change at all the structure of the algorithm.
Nevertheless, the question of the robustness of the threshold evaluation (over which the tail behavior is described with a GPD),  depending on the choice of the other components, is a topic to further investigate.

It should be emphasized that determining in an unsupervised way the threshold over which we have extremes, requires, in our method, to have information before the threshold.  A natural question could be to find the minimum information required to determine the neighbor distribution of the GPD to obtain a robust estimation for the tail threshold and the GPD parameters estimation, if willing to focus on the tail only, as EVT does.

Finally, we plan to tackle the analytical study of the convergence rate of the algorithm as a function of the sample size.

Note that a R package should  appear soon online. Meantime, the R codes are available upon request.

\section*{Acknowledgments}
The first two authors acknowledge the support from the European Union's Seventh Framework Programme for research, technological development and demonstration under grant agreement no 318984 - RARE. They also warmly thank Prof. Richard Smith for insightful comments on the manuscript.

\vspace{-1ex}
\bibliographystyle{apalike}


\newcommand{\noopsort}[1]{}

\newpage

\appendix

{\Large\bf APPENDIX}

\section{STUDY OF THE ALGORITHM CONVERGENCE}
\label{sec:App1}

As already commented, the algorithm convergence does not depend on the number \linebreak[4] ($\ge 2$) of components. Therefore, we develop its analysis when considering two components (\emph{i.e.} $u_1=u_2$; no exponential component), a Gaussian distribution and a GPD,  with a same  weight  to each one. In the following, we denote by $u$ the junction point linking those two distributions. We mention that for this  hybrid model, named G-GPD, the constraint $\beta=\xi u$ can be relaxed. The  parameters vector of the G-GPD model is $\boldsymbol{\theta}=[\mu,\sigma,u]$. The two-component algorithm (see \cite{mlsp13_hal,Debbabi2014a}) estimates the parameters $\boldsymbol{p}=[\mu,\sigma]$ and $u$ alternatively. Let us give its pseudo-code for more clarity.
\begin{algorithm}[H]
\caption{Iterative and unsupervised algorithm for  the G-GPD parameters estimation}
\label{algo1}
\begin{algorithmic}[1]
\STATE  Initialization of $\widetilde{u}^{(0)}$, $\varepsilon >0$, and $k_{max}$.\\
\STATE Determination of the empirical cdf $H_n$ associated with the  sample $\boldsymbol{x}=(x_i)_{1\leq i \leq n}$.\\
\STATE Iterative process:
\begin{itemize}
\item $k \leftarrow 1$
\begin{itemize}
 \item[Step 1 -] Estimation  of $\boldsymbol{\widetilde{p}^{(k)}}=[\widetilde{\mu}^{(k)},\widetilde{\sigma}^{(k)}] $:

$$\displaystyle \boldsymbol{\widetilde{p}^{(k)}}\leftarrow  \underset{p \in \mathcal{D}_p}{argmin}\left\| H(x;\boldsymbol{\theta}\mid{\widetilde{u}^{(k-1)}})-H_n(x)\right\|_2^2,$$\\

where $\boldsymbol{\theta}\mid{\widetilde{u}^{(k-1)}}$ represents $\boldsymbol{\theta}$ for a fixed $u=\widetilde{u}^{(k-1)}$, and $\mathcal{D}_p$ is the domain of $\boldsymbol{p}$ for $\boldsymbol{x}$.\\

\item[Step 2 -] Estimation  of $\widetilde{u}^{(k)}$:
$$\displaystyle \widetilde{u}^{(k)}\leftarrow \underset{u \in \mathcal{D}_u}{argmin}\left\| H(x;\boldsymbol{\theta}\mid{\boldsymbol{\widetilde{p}^{(k)}}})-H_n(x)\right\|_2^2,$$\\

here $\boldsymbol{\theta}\mid{\boldsymbol{\widetilde{p}^{(k)}}}$ means $\boldsymbol{\theta}$ for $\boldsymbol{p}=\boldsymbol{\widetilde{p}^{(k)}}$, and $\mathcal{D}_u$ is the $u$ domain according to $\boldsymbol{x}$.\\

\end{itemize}
 \item $k \leftarrow k+1$\\

 until {$(|\widetilde{u}^{(k)}-\widetilde{u}^{(k-1)}|< \varepsilon)$ or $(k=k_{max})$}.\\
 \end{itemize}
\STATE Return $\boldsymbol{\theta^{(k)}}=\big[\widetilde{u}^{(k)},\widetilde{\mu}^{(k)},\widetilde{\sigma}^{(k)}\big].$
\end{algorithmic}
\end{algorithm}

The convergence study is in two main steps. The first one, given in Appendix \ref{sec:App1}, consists of the analytical proof of the existence of stationary points. The algorithm, which consists of a sequence of minimization, does not rely on the optimization of a cost function
by seeking a trajectory to reach an extremum of an error surface. As a consequence, the  existence of a stationary point is not guaranteed, neither the convergence towards it; it has to be proved.
The second step consists in checking that the algorithm converges to a unique stationary point. It is done numerically, performing various simulations changing each time the initialization (see Appendix \ref{ssec:simul2}). We observe that, whatever the initialization, the algorithm converges to the same stationary point. The analytical proof of this second step is still an open problem.
\vspace{-2ex}

\subsection{Existence of Stationary Points}
\label{ssec:proof}

\vspace{-2ex}
First, let us present the theoretical framework in which the existence of stationary points  has been proved. For a given realization $\boldsymbol{x}=(x_i)_{1\leq i \leq n}$ and given parameters $\kappa, \tau \in\{\boldsymbol{p},u\}$ with $\kappa\neq\tau$,
we consider the  function:
\begin{eqnarray*}
  \varphi_\kappa : \mathcal{D}_{\tau} &\rightarrow& \mathcal{D}_\kappa \\
  \tau &\mapsto& \varphi_\kappa(\tau;\boldsymbol{x})=\underset{\kappa\in\mathcal{D}_\kappa}{argmin}\;\textstyle \So_{\tau}(\kappa;\boldsymbol{x}),
\end{eqnarray*}
where for $\boldsymbol{\theta}$ given $\tau$ (denoted by $\boldsymbol{\theta}\mid \tau$), $\textstyle \So_{\tau}$ is defined by:
\begin{eqnarray*}
\textstyle \So_{\tau}: \mathcal{D}_\kappa &\rightarrow& \R
 \\
  \kappa &\mapsto& \textstyle \So_{\tau}(\kappa;\boldsymbol{x})=\displaystyle \sum_{i=1}^n \big(H(x_i;\boldsymbol{\theta}\mid \tau)-H_n(x_i)\big)^2
\end{eqnarray*}
\vspace{-2ex}\\
To check that $\varphi_\kappa$ is a  mapping, it is enough to show that $\So_{\tau}$ admits a unique minimum, for any $\tau\in\{\boldsymbol{p},u\}$, with $\boldsymbol{p}=[\mu,\sigma]\in\R\times \R_+^*$ and $u\in\R_+$. To do so, fixing $\tau$, for instance $\tau=\boldsymbol{p}$ (it would be the same for $\tau=u$), we show that $\So_p(u;\boldsymbol{x})$ is a strongly quasiconvex function w.r.t. $u$, which is deduced from the strict convexity w.r.t. $u$ of the function $H(.;\boldsymbol{\theta}\mid \boldsymbol{p})$,  denoted by $H(.;u)$ for simplicity. Indeed we can write:
$\forall u_1, u_2\in \mathcal{D}_u$, with $u_1<u_2$, $\lambda\in[0,1]$, and $\forall i\in\{1,\cdots,n\}$:
$\displaystyle  H(x_i;\lambda u_1+(1-\lambda)u_2)<H(x_i;u_2)$,
which implies:\\
{\small $\displaystyle  \So_p(\lambda u_1+(1-\lambda)u_2;\boldsymbol{x})\hspace{-0.1cm}=\hspace{-0.1cm}\sum_{i=1}^n\hspace{-0.1cm} \big(H(x_i;u_1+(1-\lambda)u_2)-H_n(x_i)\big)^2\hspace{-0.1cm}<\hspace{-0.1cm}\sum_{i=1}^n\hspace{-0.1cm} \big(H(x_i;u_2)-H_n(x_i)\big)^2\hspace{-0.1cm}=\hspace{-0.1cm}\So_p(u_2;\boldsymbol{x}).$}\\
Hence, $\forall u_1,\;u_2\;\in \mathcal{D}_u$, $u_1<u_2$, and  $\lambda\in[0,1]$, {\small $\So_p(\lambda u_1+(1-\lambda)u_2;\boldsymbol{x}) < \max(\So_p(u_1;\boldsymbol{x}),\So_p(u_2;\boldsymbol{x}))$}. Hence, $\So_p(u;\boldsymbol{x})$ is  strongly quasiconvex on $\mathcal{D}_u$, a compact of $\mathbbm{R}$, which ensures that it admits a unique minimum. The  strict convexity w.r.t. $u$ of $H$ follows from tedious computations to show that the second derivative of $H$  w.r.t. $u$ is  positive.\\

Using $\displaystyle \varphi_u: \mathcal{D}_p \,\rightarrow\, \mathcal{D}_u$ and $\displaystyle \varphi_p: \mathcal{D}_u \,\rightarrow\, \mathcal{D}_p$, the two steps of the first iteration of the algorithm  can be given,  for a fixed $\widetilde{u}^{(0)}$,  by the following relations:
\begin{equation}
\nonumber
\left\{
  \begin{array}{ll}
     \boldsymbol{\widetilde{p}^{(1)}} =\varphi_p(\widetilde{u}^{(0)};\boldsymbol{x}), \\
      \widetilde{u}^{(1)} = \varphi_u(\widetilde{p}^{(1)};\boldsymbol{x})=\varphi_u(\varphi_p(\widetilde{u}^{(0)};\boldsymbol{x});\boldsymbol{x}).
  \end{array}
\right.
\end{equation}
More generally, for any $k\ge 1$, we can write
\begin{equation}
\label{R3}
\widetilde{u}^{(k)} = \phi(\widetilde{u}^{(k-1)};\boldsymbol{x}),
\end{equation}
where the function $\phi$ is defined from $\mathcal{D}_u$ to $\mathcal{D}_u$ by:
$\displaystyle  \phi(u;\boldsymbol{x})=\varphi_u(\varphi_p(u;\boldsymbol{x});\boldsymbol{x})$.

Consequently,  the algorithm  can also be expressed as:
\begin{center}
\begin{algorithm}[H]
\caption{Iterative and unsupervised algorithm for  the G-GPD parameters estimation \\(version 2)}
\label{algo2}
\begin{algorithmic}[1]
\STATE  Initialization of $\widetilde{u}^{(0)}$, $\varepsilon>0$, and $k_{max}$.\\
\STATE Determination of the empirical cdf $H_n$ according to $\boldsymbol{x}$.\\

\STATE Iterative process:

 $k \leftarrow 1$
 $$\displaystyle \widetilde{u}^{(k)}=\phi(\widetilde{u}^{(k-1)};\boldsymbol{x})$$
  $k \leftarrow k+1$\\
until {$(|\widetilde{u}^{(k)}-\widetilde{u}^{(k-1)}|< \varepsilon)$ or $(k=k_{max})$}.\\
\STATE Return $\widetilde{u}^{(k)}.$
\end{algorithmic}
\end{algorithm}
\end{center}
\vspace{-1ex}
A way to prove the existence of stationary points of Algorithm \ref{algo2} is to demonstrate the existence of fixed-points of the function $\phi$. To do so, we build on the fixed-point theorem. Several versions of this theorem exist in the literature \emph{e.g.} the version of  Banach  (see \cite{Banach2007}), or of Markov-Kakutani (see \cite{Markov1993}), or of Schauder  (see \cite{Schauder2009}), or of Brouwer (see \cite{Brouwer1941}). In this work, we consider the latter one, as its hypotheses are, in our case, more straightforward to check. This theorem states that {\it every continuous function from a closed ball of a Euclidean space into itself has a fixed point}.
It implies that the functional  $\phi$ admits at least one fixed point if the following two conditions, ($\mathcal{C}_1$) and ($\mathcal{C}_2$), are satisfied:
\begin{description}
  \item[($\mathcal{C}_1$) :] $\mathcal{D}_u$ is a closed ball of a Euclidean space.

  \item[($\mathcal{C}_2$) :] $\phi$ is continuous on $\mathcal{D}_u$.
\end{description}
The conditions  ($\mathcal{C}_1$) is clearly satisfied: for a realization $x$,  $\mathcal{D}_u=[0,max(\boldsymbol{x})]$ is a closed ball of $\R$ that is a Euclidian space.

Now, to verify  ($\mathcal{C}_2$), we prove that  $\varphi_u$ and $\varphi_p$ are both continuous on their domains (since $\phi$ is the composite function:  $\phi=\varphi_u\circ\varphi_p$) using the Heine-Cantor theorem and the Ramsay et al.'s one that we recall here.

{\it \textsc{Theorem \cite{Ramsay2007}}\\
Let $\mathcal{X}$ and $\mathcal{Y}$ be metric spaces with $\mathcal{X}$ closed and bounded. Let {\small\begin{eqnarray*}
g: \mathcal{X}\times\mathcal{Y} &\rightarrow& \R \\
  (x,\alpha) &\rightarrow& g(x,\alpha)
\end{eqnarray*}
}
\vspace{-.5cm}\\
 be uniformly continuous in $x$ and $\alpha$, such that  $\displaystyle x(\alpha)=\underset{x\in\mathcal{X}}{argmin}\,g(x,\alpha)$
is well defined for all $\alpha \in \mathcal{Y}$. Then the function $x(\alpha):\mathcal{Y}\rightarrow\mathcal{X}$ is continuous.
}

The proof of the continuity of the two functions $\varphi_p$ and $\varphi_u$ being the same, let us consider for instance the function $\varphi_p$.
Using Ramsay  et al.'s theorem, we need to check that $\mathcal{D}_p$ is a compact  and that $\textstyle \So_u$ is uniformly continuous on $\mathcal{D}_p$, to conclude to the continuity of $\varphi_p$. The first condition, $\mathcal{D}_p$ is a compact of $\R^2$, is satisfied when noticing that we are working with a Gaussian density, with finite mean  and variance, hence which is bounded.

Now, as $\mathcal{D}_p$ is a compact,  it is sufficient to  show that  $\textstyle \So_u$ is continuous on $\mathcal{D}_p$ to deduce,  by the Heine-Cantor theorem, its uniform continuity. To do so,  we just need to study the continuity of $H$  w.r.t. $\boldsymbol{p}$ to deduce the continuity of $\So_u$ w.r.t. $\boldsymbol{p}$, since  $\textstyle \So_u(\boldsymbol{p};\boldsymbol{x})=\displaystyle \sum_{i=1}^n \big(H(x_i;\boldsymbol{\theta}\mid u)-H_n(x_i)\big)^2$. We recall here that, by construction, $H$ is continuous w.r.t. $x$ and not to its parameters. Hence, its continuity according to $\boldsymbol{p}$ remains to be proved.  Since $H$ is composed of two functions (see \eqref{cdf} for $u_1=u_2$), the Gaussian cdf and the GPD, we will study the continuity of each one w.r.t. $\boldsymbol{p}$. The continuity of  the Gaussian cdf $F$ as a function of $\boldsymbol{p}$ is obvious since it means to look at the  continuity of its likelihood w.r.t. $\boldsymbol{p}=[\mu,\sigma]$. Now, for  the GPD  $G$, its parameters  $\xi$ and $\beta$  are expressed as fonctions of $\boldsymbol{p}$:
$\displaystyle  \beta(\boldsymbol{p})= 1/f(u;\boldsymbol{p}),\;\;\mbox{and}\;\; \xi(\boldsymbol{p})=-1+(u-\mu)\beta(\boldsymbol{p})/\sigma^2$,
and are both continuous in $\boldsymbol{p}$. Hence $G$ is continuous in  $\boldsymbol{p}$ as the composition of continuous functions w.r.t. $\boldsymbol{p}$.

Finally, we can deduce the continuity of  the function $\textstyle \So_u$ on $\mathcal{D}_p$ as a composition, sum and products, of continuous functions on $\mathcal{D}_p$, from which we conclude to the continuity  of $\varphi_p$ on $\mathcal{D}_u$.

{\it Conclusion:} The functional  $\phi$ is continuous on  $\mathcal{D}_u$ as a composition of two continuous functions: $\varphi_p$ and $\varphi_u$. Hence the existence of at least one fixed-point according to  the Brouwer fixed-point theorem. Consequently, the algorithm  admits at least one stationary point.
Since the method does not follow a path on an error surface, it is free from local minima traps, as are the standard
gradient search based methods. In the next section, we perform simulations to check if  the algorithm converges to a unique stationary point regardless to its initialization.
\vspace{-1cm}
\subsection{Numerical Study of the Algorithm Convergence}
\label{ssec:simul2}

\vspace{-2ex}

To study numerically the convergence of the algorithm to a unique attractive stationary point, we consider the recurrent sequence  $\{\widetilde{u}^{(k+1)}=\phi(\widetilde{u}^{(k)})\}_{k\in\N^*}$,  obtained when applying Algorithm \ref{algo1} on a generated G-GPD distributed data with a fixed parameter $\boldsymbol{\theta}$. Different initial  values  of this sequence are considered; they have been selected in the interval  $\displaystyle I=[q_{_{25\%}},q_{_{50\%}}]$, since we assumed uniform weights for the two components (see \cite{Debbabi2015b}, Remark in Section 2).
For illustration, we report here two examples among all those performed to test the convergence.

\begin{figure}[H]
  \centering
  \includegraphics[scale=0.35]{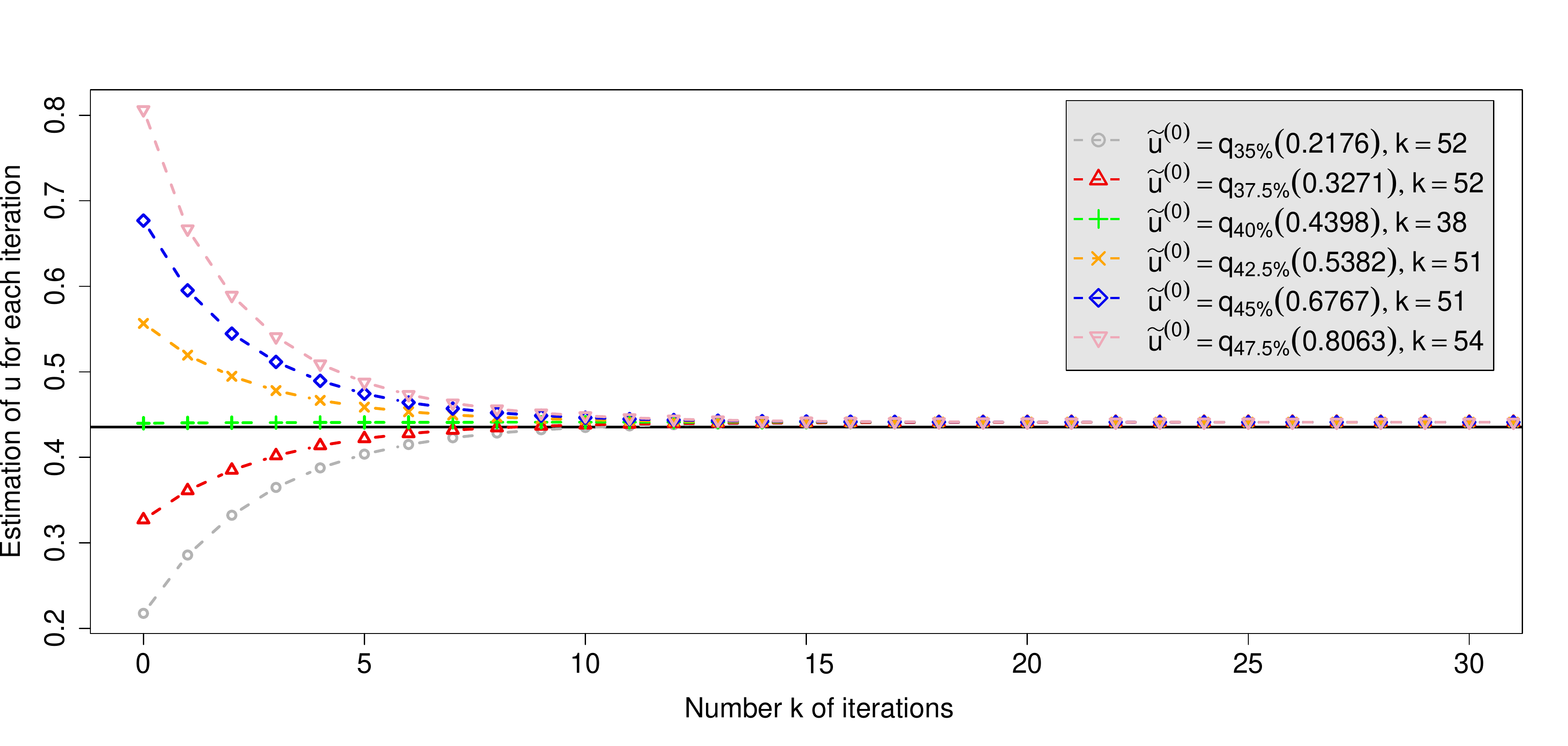}
\vspace{-2ex}
  \caption{Study of the convergence of the recurrent sequence $\{\widetilde{u}^{(k+1)}=\phi(\widetilde{u}^{(k)})\}_{k\in\N^*}$  regarding the initial value  $\widetilde{u}^{(0)}$. Example $1$ for $\boldsymbol{\theta}=[0,1,0.4354]$ and $u^*=0.4354=q_{_{39.42\%}}$.}\label{CV1}
\end{figure}

Example 1. For $\boldsymbol{\theta}=[0,1,0.4354]$ with $\xi=0.2$ and $\beta=2.7558$, we present in Figure \ref{CV1} the corresponding recurrent sequence  $\{\widetilde{u}^{(k+1)}=\phi(\widetilde{u}^{(k)})\}_{k\in\N^*}$, where  the initial value \linebreak[4] $\widetilde{u}^{(0)}\in\{q_{_{35\%}},q_{_{37.5\%}},q_{_{40\%}},q_{_{42.5\%}},q_{_{45\%}},q_{_{47.5\%}}\}$.
As shown in this figure, regardless the choice of $\widetilde{u}^{(0)}$ in $I$, the algorithm converges to the fixed value of $u=0.4354$ (represented by a continuous horizontal line), denoted by $u^*$. We observe that:\\
\begin{enumerate}
  \item  If $\widetilde{u}^{(0)}< u^{*}$, the associated recurrent sequence is non decreasing, as for instance for the gray cercles curve with $\widetilde{u}^{(0)}=q_{_{35\%}}$ and the red triangles (upwards oriented) one with  $\widetilde{u}^{(0)}=q_{_{37.5\%}}$;\\
  \item If $\widetilde{u}^{(0)}> u^{*}$,  the associated recurrent sequence is non increasing, \emph{e.g.} the blue diamonds curve  for $\widetilde{u}^{(0)}=q_{_{45\%}}$ and the pink triangles (downwards oriented) curve for  $\widetilde{u}^{(0)}=q_{_{47.5\%}}$ .\\
 \end{enumerate}

Consequently, based on Figure \ref{CV1}, regardless the choice of  $\widetilde{u}^{(0)}\in I$, the recurrent sequence \linebreak[4] $\{\widetilde{u}^{(k+1)}=\phi(\widetilde{u}^{(k)})\}_{k\in\N^*}$ is  monotone on $\mathcal{D}_u$ and  converges to a unique attractive  stationary point that corresponds to $u^*$.

\begin{figure}[H]
  \centering
  \includegraphics[scale=0.35]{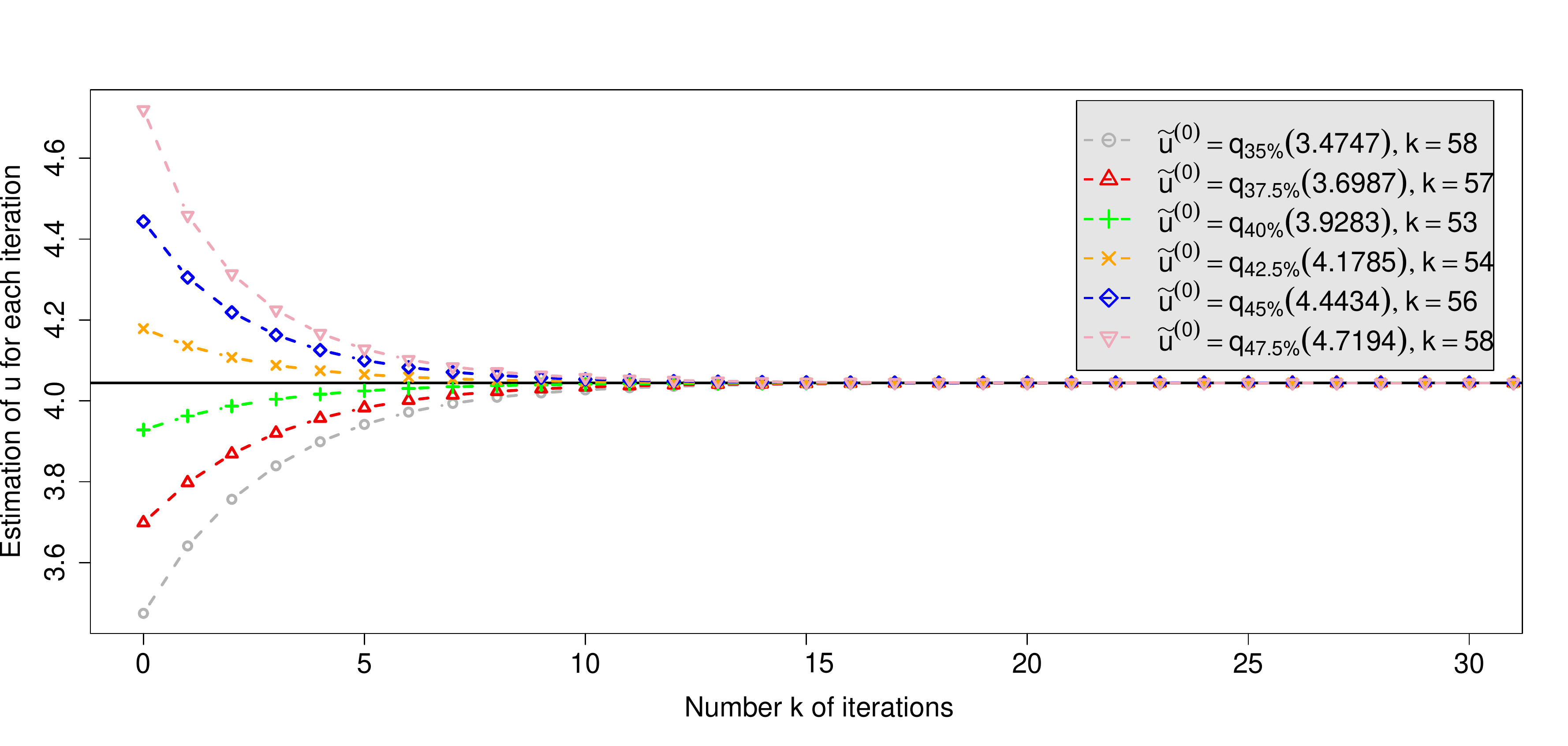}
\vspace{-2ex}
  \caption{Study of the convergence of the recurrent sequence $\{\widetilde{u}^{(k+1)}=\phi(\widetilde{u}^{(k)})\}_{k\in\N^*}$  regarding the initial value  $\widetilde{u}^{(0)}$. Example $2$ for $\boldsymbol{\theta}=[3,2,4.0443]$ and  $u^*=4.0443=q_{_{41.18\%}}$}\label{CV2}
\end{figure}

Example 2. Consider now for $\boldsymbol{\theta}=[3,2,4.0443]$ with $\xi=0.5$ and $\beta=5.7454$. It is illustrated in Figure \ref{CV2} and leads to the same observations.

 An additional remark concerns the number of iterations. We could observe in the simulation study that the closest to $u^*$ is $\widetilde{u}^{(0)}$, the fastest is the convergence, as expected. It appears clearly on the  two examples (see the green '$+$' marks curve in both figures).

%
\vspace{-2ex}
\subsection{Extension to the three-component case}
\label{ssec:3comp}

Extending the convergence study of the algorithm to three components is straightforward
and follows the same logic as for two components.

The difference between the two- and three-component algorithms concerns only the definition of the parameters vector (taking more components implies having more parameters; we also assumed in this paper $\xi$ to be positive, considering a heavy tailed distribution belonging to the Fr\'{e}chet maximum domain of attraction), the data scale (to have more points in the tail) and the stop condition (that we improved, formulating it in terms of the goodness-of-fit of both the entire distribution and the tail distribution). The estimation of the vector parameter $\boldsymbol{\theta}=[\mu,\sigma,u_2,\xi]$ of the G-E-GPD algorithm is  broken down into two parts, the estimation of $\boldsymbol{p}=[\mu,\sigma,u_2]$ and of $\xi$, alternately (as for the two-component algorithm). The algorithm can then be represented by  a functional of $\xi$; it is summarized in the pseudo-code below (see Algorithm \ref{algo3}).

To conclude, the modifications we introduced have been chosen to enhance the parameters estimation, when adapting the algorithm to a larger number of parameters. It does not interfere in the functional principle of the algorithm, nor on how to study its convergence.

\begin{algorithm}[H]
\caption{Iterative and unsupervised algorithm for  the G-E-GPD parameters estimation}
\label{algo3}
\begin{algorithmic}[1]
\STATE Initialization of $\boldsymbol{\widetilde{p}^{(0)}}$, $\alpha$, $\varepsilon>0$, and  $k_{max}$, then  initialization of $\widetilde{\xi}^{(0)}$:\\
$$\displaystyle \widetilde{\xi}^{(0)} \leftarrow \underset{\xi>0}{argmin}\left\| H(\boldsymbol{y};\boldsymbol{\theta}\mid{\boldsymbol{\widetilde{p}^{(0)}}})-H_n(\boldsymbol{y})\right\|_2^2,$$\\
where  $H_n$ is the empirical cdf of X. We note that this distance is computed on the points $\boldsymbol{y}=(y_j)_{1\leq j \leq m}$ defined in \eqref{newdata}.\\
\STATE Iterative process:
\begin{itemize}
\item $k \leftarrow 1$
\begin{itemize}
 \item[Step 1 -] Estimation of $\boldsymbol{\widetilde{p}^{(k)}}$:
$$\displaystyle \boldsymbol{\widetilde{p}^{(k)}} \leftarrow \underset{u_2 \in\R_+}{\underset{(\mu,\sigma) \in\R\times\R^*_+}{argmin}}\left\| H(\boldsymbol{y};\boldsymbol{\theta}\mid{\widetilde{\xi}^{(k-1)}})-H_n(\boldsymbol{y})\right\|_2^2$$\\
  \item[Step 2 -] Estimation of  $\widetilde{\xi}^{(k)}$:
$$
\widetilde{\xi}^{(k)} \leftarrow \underset{\xi>0}{argmin}\left\| H(\boldsymbol{y};\boldsymbol{\theta}\mid{\boldsymbol{\widetilde{p}^{(k)}}})-H_n(\boldsymbol{y})\right\|_2^2,
$$
\end{itemize}
\item $k \leftarrow k+1$\\
 until {\small $\Big( d(H(\boldsymbol{y};\boldsymbol{\theta^{(k)}}),H_n(\boldsymbol{y}))< \varepsilon \;\;\mbox{and}\;\; d(H(\boldsymbol{y_{q_{_\alpha}}};\boldsymbol{\theta^{(k)}}),H_n(\boldsymbol{y_{q_{_\alpha}}})) < \varepsilon \Big)$} or $\big(k=k_{max}\big)$.
\end{itemize}
\STATE Return $\boldsymbol{\theta^{(k)}}=\big[\widetilde{\mu}^{(k)}, \widetilde{\sigma}^{(k)}, \widetilde{u}_2^{(k)},\widetilde{\xi}^{(k)}\big].$
\end{algorithmic}
\end{algorithm}


\section{ADDITIONAL MONTE-CARLO SIMULATIONS RESULTS OF THE \\G-E-GPD ALGORITHM}\label{sec:App2}
\vspace{-.5cm}

For an accurate interpretation of MC simulations results, we complete Section~\ref{sec:Simul} of the paper with other  examples, varying the various parameters $\boldsymbol{\theta}=[\mu,\sigma,u_2,\xi]$ and the sample size $n$. For all examples, we fix the risk $\delta=5\%$, the test set size $l=n$, and the order of quantile (used in Condition C2) $\alpha=0.8$.

Table~\ref{tab11}  illustrates the case when $\boldsymbol{\theta}=[1,1,12,0.5]$ and $n\in\{10^3,10^4,10^5\}$, to compare the results with Table~\ref{tab1} and to test the robustness of the tail index estimation when having a higher threshold ($u_2=12=q_{92.81\%}$) than in Table~\ref{tab1}.

In Table \ref{tab11bis}, we fix $\mu$, $\sigma$, and $\xi$, but vary the thresholds $u_1$ and $u_2$ (and also $n$, with $n\in\{10^4,10^5\}$), to observe the variability range for the parameters estimations.\\
\renewcommand\arraystretch{.8}
\begin{table}[H]
\caption{$\boldsymbol{\theta}=[1,1,12,0.5]$, $\rho=0.9$, with various $n$} \label{tab11}
\centering
\scalebox{.8}{
\begin{tabular}{|c|c|c|c|c|c|}
\cline{4-6}
\multicolumn{3}{c|}{ } & $n=10^3$ & $n=10^4$ & $n=10^5$ \\ \hline
\multirow{8}{*}{\vspace{-5cm}\rotatebox{90}{Parameters}}&\multirow{4}{*}{$\mu=1$ } & $\widetilde{\mu}$ & $1.0358$ & $1.0057$ & $0.9989$\\  \cline{3-6}
&  & $\tilde{S}_N^{^\mu}$ & $3.74$ $10^{-2}$ & $3.38$ $10^{-3}$ & $3.34$ $10^{-4}$ \\  \cline{3-6}
&  & MSE$_{\mu}$ & $3.83$ $10^{-2}$ & $3.38$ $10^{-3}$ & $3.32$ $10^{-4}$ \\  \cline{3-6}
\multirow{8}{*}{}&\multirow{4}{*}{$\sigma=1$ } & $p_{T_{\widetilde{\mu},N}}$ & $0.8529$ & $0.9207$ & $ 0.9539$ \\  \cline{2-6}
 & & $\widetilde{\sigma}$ & $1.0236$  & $ 1.0013$ & $0.9987$ \\   \cline{3-6}
 &  & $\tilde{S}_N^{^\sigma}$ & $2.16$ $10^{-2}$ & $1.66$ $10^{-3}$ & $1.71$ $10^{-4}$ \\  \cline{3-6}
 & & MSE$_{\sigma}$ & $2.2$ $10^{-2}$ & $1.65$ $10^{-3}$& $1.71$ $10^{-4}$ \\  \cline{3-6}
\multirow{8}{*}{}&\multirow{4}{*}{$u_2=12=q_{_{92.81\%}}$ } &  $p_{T_{\widetilde{\sigma},N}}$ & $ 0.8722$ & $0.974$ & $ 0.9232$\\  \cline{2-6}
 & & $\widetilde{u_2}$ & $11.8886$ & $12.0062$  & $12.0048$\\   \cline{3-6}
 &  & $\tilde{S}_N^{^{u_2}}$ & $2.0846$ & $1.40$ $10^{-1}$ & $3.85$ $10^{-2}$ \\  \cline{3-6}
 & ($u_1=1.25=q_{_{27.23\%}}$) & MSE$_{u_2}$ & $2.0762$& $1.95$ $10^{-1}$ & $1.59$ $10^{-2}$\\  \cline{3-6}
\multirow{8}{*}{}&\multirow{4}{*}{$\xi=0.5$ } & $p_{T_{\widetilde{u_2},N}}$ & $ 0.9385$ & $ 0.9888$ & $0.9692$\\  \cline{2-6}
 & & $\widetilde{\xi}$& $ 0.5074$ & $0.4994$ & $0.4999$ \\   \cline{3-6}
  &  & $\tilde{S}_N^{^\xi}$ & $3.84$ $10^{-3}$ & $3.44$ $10^{-4}$ & $3.07$ $10^{-5}$ \\  \cline{3-6}
 & & MSE$_{\xi}$ & $3.85$ $10^{-3}$& $3.41$ $10^{-4}$& $3.04$ $10^{-5}$ \\  \cline{3-6}
 & & $p_{T_{\widetilde{\xi},N}}$ & $0.9042$ & $ 0.9749$& $ 0.9971$\\   \hline
 \multicolumn{3}{|c|}{ Average  execution time (seconds)} &$8.52$  & $25.53$  & $208.46$ \\ \hline
  \multicolumn{3}{|c|}{ Average  iterations number} &$92$ &$102$ &$110$ \\ \hline
 \multicolumn{3}{|c|}{ $\mathcal{D}$} &$3.15$ $10^{- 3}$& $3.39$ $10^{- 4}$ & $3.07$ $10^{-5}$\\ \hline
\end{tabular}
}
\end{table}
\vspace{-1cm}
\renewcommand\arraystretch{.8}
\begin{table}[H]
\caption{$\mu=\sigma=1$, $\xi=0.3$, $u_2\in\{2.7,3,12,12\}$, and $\rho=0.9$} \label{tab11bis}
\vspace{.5cm}
\centering
\scalebox{.75}{
\begin{tabular}{|c|c|c|c|c|c|c|}
\cline{4-7}
\multicolumn{3}{c|}{ } & $u_2=2.7=q_{_{92.58\%}}$ & $u_2=3=q_{_{94.52\%}}$ & $u_2=12=q_{_{98.28\%}}$ & $u_2=12=q_{_{98.28\%}}$\\
\multicolumn{3}{c|}{ } & ($u_1=2.6=q_{_{91.64\%}}$)& ($u_1=2.44=q_{_{89.34\%}}$) & ($u_1=1.36=q_{_{38.09\%}}$) & ($u_1=1.36=q_{_{38.09\%}}$)\\
\multicolumn{3}{c|}{ } & $l=n=10^4$ & $l=n=10^4$ & $l=n=10^4$ & $l=n=10^5$\\ \hline
\multirow{8}{*}{\vspace{-5cm}\rotatebox{90}{Parameters}}&\multirow{4}{*}{$\mu=1$ } & $\widetilde{\mu}$  & $0.9991$ & $0.9996$ & $1.0066$ & $0.9989$\\  \cline{3-7}
&  & $\tilde{S}_N^{^\mu}$  & $1.43$ $10^{-4}$ & $1.89$ $10^{-4}$ & $1.59$ $10^{-3}$ & $2.05$ $10^{-4}$\\  \cline{3-7}
&  & MSE$_{\mu}$  & $1.42$ $10^{-4}$ & $1.87$ $10^{-4}$ & $1.62$ $10^{-3}$& $2.04$ $10^{-4}$ \\  \cline{3-7}
\multirow{8}{*}{}&\multirow{4}{*}{$\sigma=1$ } & $p_{T_{\widetilde{\mu},N}}$  & $0.941$ & $ 0.9801$ & $ 0.867$ & $0.9412$\\  \cline{2-7}
 & & $\widetilde{\sigma}$   & $ 0.9986$ & $0.9996$ & $1.0034$ & $0.9994$\\   \cline{3-7}
 &  & $\tilde{S}_N^{^\sigma}$  & $1.21$ $10^{-4}$ & $1.27$ $10^{-4}$& $8.57$ $10^{-4}$ & $1.33$ $10^{-4}$\\  \cline{3-7}
 & & MSE$_{\sigma}$  & $1.22$ $10^{-4}$& $1.26$ $10^{-4}$ & $8.6$ $10^{-4}$ & $1.32$ $10^{-4}$\\  \cline{3-7}
\multirow{8}{*}{}&\multirow{4}{*}{$u_2$ } &  $p_{T_{\widetilde{\sigma},N}}$  & $0.9058$  & $ 0.9763$ & $0.9054$ & $0.9626$\\  \cline{2-7}
 & & $\widetilde{u_2}$  & $2.868$  & $2.9467$& $12.5296$ & $12.0699$\\   \cline{3-7}
 &  & $\tilde{S}_N^{^{u_2}}$  & $9.32$ $10^{-2}$ & $1.08$ $10^{-1}$ & $5.1441$ & $1.71$ $10^{-1}$\\  \cline{3-7}
 & & MSE$_{u_2}$ & $1.2$ $10^{-1}$ & $1.09$ $10^{-1}$& $5.3732$ & $1.74$ $10^{-1}$\\  \cline{3-7}
\multirow{8}{*}{}&\multirow{4}{*}{$\xi=0.3$ } & $p_{T_{\widetilde{u_2},N}}$  & $0.582$ & $0.8713$& $0.8623$ &$0.8153$ \\  \cline{2-7}
 & & $\widetilde{\xi}$  & $0.2936$ & $0.2974$& $0.2972$ & $0.2925$\\   \cline{3-7}
  &  & $\tilde{S}_N^{^\xi}$ & $1.73$ $10^{-4}$ & $1.93$ $10^{-4}$ & $1.87$ $10^{-3}$ & $1.37$ $10^{-4}$\\  \cline{3-7}
 & & MSE$_{\xi}$ & $2.11$ $10^{-4}$& $1.98$ $10^{-4}$ & $1.91$ $10^{-3}$ & $1.38$ $10^{-4}$\\  \cline{3-7}
 & & $p_{T_{\widetilde{\xi},N}}$ &$0.6301$ & $ 0.8538$& $ 0.864$ & $0.896$\\   \hline
 \multicolumn{3}{|c|}{ Average  execution time (seconds)}   & $37.66$  & $30.87$ & $135.63$ & $1449.06$\\ \hline
  \multicolumn{3}{|c|}{ Average  iterations number} &$107$ &$86$ & $510$  & $543$\\ \hline
 \multicolumn{3}{|c|}{ $\mathcal{D}$} & $2.59$ $10^{- 4}$ & $2.52$ $10^{-4}$ & $5.04$ $10^{-4}$ & $4.41$ $10^{-5}$\\ \hline
\end{tabular}
}
\end{table}

From now on, we fix the size $n$ choosing e.g. $n=10^4$, and vary other parameters.
In Tables \ref{tab22} and \ref{tab22bis}, we vary the value of the tail index $\xi$, choosing the parameters $\mu$ and $\sigma$ of the Gaussian component as $\mu=\sigma=2$ (instead of 1 as in the previous tables).\\

\begin{minipage}{10.5cm}
\begin{table}[H]
\caption{$\mu=\sigma=2$, $\xi=0.5$, $u_2\in\{5,8,12\}$, $\rho=0.9$, and $n=10^4$} \label{tab22}
\vspace{0.5cm}
\centering
\scalebox{0.65}{
\begin{tabular}{|c|c|c|c|c|c|}
\cline{4-6}
\multicolumn{3}{c|}{ } & $u_2=5=q_{_{84.64\%}}$ & $u_2=8=q_{_{90.8\%}}$ & $u_2=12=q_{_{92.22\%}}$ \\
\multicolumn{3}{c|}{ } & $(u_1=4.4=q_{_{80.21\%}}$) & ($u_1=3.5=q_{_{63.8\%}}$) & ($u_1=3=q_{_{48.26\%}}$) \\ \hline
\multirow{8}{*}{\vspace{-5cm}\rotatebox{90}{Parameters}}&\multirow{4}{*}{$\mu=2$ } & $\widetilde{\mu}$ & $1.992$ & $1.9972$ & $2.0061$\\  \cline{3-6}
&  & $\tilde{S}_N^{^\mu}$ & $8.69$ $10^{-4}$ & $1.64$ $10^{-3}$ & $3.35$ $10^{-3}$ \\  \cline{3-6}
&  & MSE$_{\mu}$ & $9.23$ $10^{-4}$ & $1.63$ $10^{-3}$ & $3.35$ $10^{-3}$ \\  \cline{3-6}
\multirow{8}{*}{}&\multirow{4}{*}{$\sigma=2$ } & $p_{T_{\widetilde{\mu},N}}$ & $0.7884$ &$0.9468$  & $0.9148$ \\  \cline{2-6}
 & & $\widetilde{\sigma}$ & $1.9991$  & $ 2.0003$ & $2.0036$ \\   \cline{3-6}
 &  & $\tilde{S}_N^{^\sigma}$ & $8.32$ $10^{-4}$ & $201$ $10^{-3}$ & $2.271$ $10^{-3}$ \\  \cline{3-6}
 & & MSE$_{\sigma}$ & $8.25$ $10^{-4}$ & $1.11$ $10^{-3}$& $2.09$ $10^{-3}$ \\  \cline{3-6}
\multirow{8}{*}{}&\multirow{4}{*}{$u_2$ } &  $p_{T_{\widetilde{\sigma},N}}$ &$ 0.9753$  & $0.9916$  & $0.9373$ \\  \cline{2-6}
 & & $\widetilde{u_2}$ & $4.9552$ & $7.9501$  & $11.9796$\\   \cline{3-6}
 &  & $\tilde{S}_N^{^{u_2}}$ & $1.86$ $10^{-1}$ & $8.85$ $10^{-2}$ & $2.21$ $10^{-1}$ \\  \cline{3-6}
 & & MSE$_{u_2}$ & $1.68$ $10^{-1}$& $9.01$ $10^{-2}$ & $2.2$ $10^{-1}$\\  \cline{3-6}
\multirow{8}{*}{}&\multirow{4}{*}{$\xi=0.5$ } & $p_{T_{\widetilde{u_2},N}}$ & $ 0.9174$  & $0.867$  & $0.9655$\\  \cline{2-6}
 & & $\widetilde{\xi}$& $ 0.4984$ & $0.5025$ & $0.4998$ \\   \cline{3-6}
  &  & $\tilde{S}_N^{^\xi}$ & $1.42$ $10^{-4}$ & $2.00$ $10^{-4}$ & $2.9$ $10^{-4}$ \\  \cline{3-6}
 & & MSE$_{\xi}$ & $1.44$ $10^{-4}$& $2.04$ $10^{-4}$& $2.87$ $10^{-4}$ \\  \cline{3-6}
 & & $p_{T_{\widetilde{\xi},N}}$ & $0.8935$  & $0.8594$ & $0.9935$ \\   \hline
 \multicolumn{3}{|c|}{ Average  execution time (seconds)} &$65.83$  & $43.46$  & $45.83$ \\ \hline
  \multicolumn{3}{|c|}{ Average  iterations number} &$114$ &$66$ &$80$ \\ \hline
 \multicolumn{3}{|c|}{ $\mathcal{D}$} &$2.95$ $10^{-4}$& $2.63$ $10^{- 4}$ & $3.11$ $10^{-4}$\\ \hline
\end{tabular}
}
\end{table}
\end{minipage}
\begin{minipage}{8cm}
\begin{table}[H]
\caption{$\boldsymbol{\theta}=[2,2,20,1]$,  $\rho=0.8$, \\ and  $n=10^4$.} \label{tab22bis}
\vspace{0.8cm}
\centering
\scalebox{0.65}{
\begin{tabular}{|c|c|c|c|}
\cline{4-4}
\hline
\multirow{8}{*}{\vspace{-5cm}\rotatebox{90}{Parameters}}&\multirow{4}{*}{$\mu=2$ } & $\widetilde{\mu}$ & $2.0194$\\  \cline{3-4}
&  & $\tilde{S}_N^{^\mu}$ & $2.77$ $10^{-2}$  \\  \cline{3-4}
&  & MSE$_{\mu}$ & $2.78$ $10^{-2}$  \\  \cline{3-4}
& & $T_{\widetilde{\mu},N}$ & $0.1166$ \\  \cline{3-4}
\multirow{8}{*}{}&\multirow{4}{*}{$\sigma=2$ } & $p_{T_{\widetilde{\mu},N}}$ & $0.9071$  \\  \cline{2-4}
 & & $\widetilde{\sigma}$ & $2.0057$ \\   \cline{3-4}
 &  & $\tilde{S}_N^{^\sigma}$ & $1.3$ $10^{-2}$ \\  \cline{3-4}
 & & MSE$_{\sigma}$ & $1.29$ $10^{-2}$\\  \cline{3-4}
\multirow{8}{*}{}&\multirow{4}{*}{$u_2=20=q_{_{76.56\%}}$ } &  $p_{T_{\widetilde{\sigma},N}}$ & $0.96$   \\  \cline{2-4}
 & & $\widetilde{u_2}$ & $19.8697$  \\   \cline{3-4}
 &  & $\tilde{S}_N^{^{u_2}}$ & $5.05$ $10^{-1}$  \\  \cline{3-4}
 &($u_1=2.4=q_{_{20.17\%}}$) & MSE$_{u_2}$ & $5.17$ $10^{-1}$\\  \cline{3-4}
\multirow{8}{*}{}&\multirow{4}{*}{$\xi=1$ } & $p_{T_{\widetilde{u_2},N}}$ & $0.8546$   \\  \cline{2-4}
 & & $\widetilde{\xi}$& $1.0051$ \\   \cline{3-4}
  &  & $\tilde{S}_N^{^\xi}$ & $7.49$ $10^{-4}$ \\  \cline{3-4}
 & & MSE$_{\xi}$ & $7.62$ $10^{-4}$ \\  \cline{3-4}
 & & $p_{T_{\widetilde{\xi},N}}$ & $0.8512$  \\   \hline
 \multicolumn{3}{|c|}{ Average  execution time (seconds)} &$20.79$  \\ \hline
  \multicolumn{3}{|c|}{ Average  iterations number} &$185$\\ \hline
 \multicolumn{3}{|c|}{ $\mathcal{D}$} &$2.86$ $10^{-4}$ \\ \hline
\end{tabular}
}
\end{table}
\end{minipage}

\vspace{5ex}
Finally, in the last two tables \ref{tab0.5} and \ref{tab05}, we fix $\mu=0$, but vary the standard deviation $\sigma$ of the Gaussian component,  the threshold $u_1$ and $u_2$, and consider two cases of tail index, when having a heavy tail ($\xi<1$) and a  very heavy tail ($\xi>1$).

\renewcommand\arraystretch{1.35}
\begin{minipage}{10cm}
\begin{table}[H]
\centering
\caption{$\mu=0$, $\sigma=0.5$, $\xi=0.4$, $u_2\in\{1,10\}$, $\rho=0.9$,\\ and $n=10^4$.} \label{tab0.5}
\vspace{0.5cm}
\scalebox{0.65}{
\begin{tabular}{|c|c|c|c|c|}
\cline{4-5}
\multicolumn{3}{c|}{ } & $u_2=1=q_{_{95.63\%}}$ & $u_2=10=q_{_{96.55\%}}$ \\
\multicolumn{3}{c|}{ } & $(u_1=0.875=q_{_{93.93\%}}$) & ($u_1=0.0875=q_{_{20.03\%}}$) \\ \hline
\multirow{8}{*}{\vspace{-5cm}\rotatebox{90}{Parameters}}&\multirow{4}{*}{$\mu=0$ } & $\widetilde{\mu}$ & $-1.00$ $10^{-3}$  & $-9.8$ $10^{-4}$\\  \cline{3-5}
&  & $\tilde{S}_N^{^\mu}$ & $4.25$ $10^{-5}$ & $1.62$ $10^{-3}$  \\  \cline{3-5}
&  & MSE$_{\mu}$ & $4.31$ $10^{-5}$ & $1.6$ $10^{-3}$  \\  \cline{3-5}
\multirow{8}{*}{}&\multirow{4}{*}{$\sigma=0.5$ } & $p_{T_{\widetilde{\mu},N}}$ &$0.8779$ & $0.9805$  \\  \cline{2-5}
 & & $\widetilde{\sigma}$ & $0.4988$  & $0.5014$ \\   \cline{3-5}
 &  & $\tilde{S}_N^{^\sigma}$ & $3.21$ $10^{-5}$ & $8.31$ $10^{-4}$  \\  \cline{3-5}
 & & MSE$_{\sigma}$ & $3.32$ $10^{-5}$ & $8.24$ $10^{-4}$\\  \cline{3-5}
\multirow{8}{*}{}&\multirow{4}{*}{$u_2$ } &  $p_{T_{\widetilde{\sigma},N}}$ & $ 0.836$ & $0.9585$   \\  \cline{2-5}
 & & $\widetilde{u_2}$ & $1.0456$ & $9.928$  \\   \cline{3-5}
 &  & $\tilde{S}_N^{^{u_2}}$ & $2.6$ $10^{-2}$ & $3.35$ $10^{-1}$ \\  \cline{3-5}
 & & MSE$_{u_2}$ & $2.78$ $10^{-2}$& $3.37$ $10^{-1}$\\  \cline{3-5}
\multirow{8}{*}{}&\multirow{4}{*}{$\xi=0.4$ } & $p_{T_{\widetilde{u_2},N}}$ & $0.7771$  & $0.901$   \\  \cline{2-5}
 & & $\widetilde{\xi}$& $ 0.3886$ & $0.4053$ \\   \cline{3-5}
  &  & $\tilde{S}_N^{^\xi}$ & $8.43$ $10^{-4}$ & $7.24$ $10^{-4}$  \\  \cline{3-5}
 & & MSE$_{\xi}$ & $9.64$ $10^{-4}$& $7.46$ $10^{-4}$ \\  \cline{3-5}
 & & $p_{T_{\widetilde{\xi},N}}$ & $0.695$ & $0.8413$  \\   \hline
 \multicolumn{3}{|c|}{ Average  execution time (seconds)} &$86.88$  & $64.02$  \\ \hline
  \multicolumn{3}{|c|}{ Average  iterations number} &$185$ &$192$ \\ \hline
 \multicolumn{3}{|c|}{ $\mathcal{D}$} &$3.45$ $10^{-4}$& $3.96$ $10^{- 4}$ \\ \hline
\end{tabular}
}
\end{table}
\end{minipage}
\begin{minipage}{8cm}
\begin{table}[H]
\centering
\caption{$\boldsymbol{\theta}=[0,5,11,1.2]$, $\rho=0.8$,\\  and  $n=10^4$.} \label{tab05}
\vspace{1.5cm}
\scalebox{0.65}{
\begin{tabular}{|c|c|c|c|}
\cline{4-4}
\hline
\multirow{8}{*}{\vspace{-5cm}\rotatebox{90}{Parameters}}&\multirow{4}{*}{$\mu=0$ } & $\widetilde{\mu}$ & $5.99$ $10^{-3}$\\  \cline{3-4}
&  & $\tilde{S}_N^{^\mu}$ & $1.42$ $10^{-2}$  \\  \cline{3-4}
&  & MSE$_{\mu}$ & $1.41$ $10^{-2}$  \\  \cline{3-4}
\multirow{8}{*}{}&\multirow{4}{*}{$\sigma=5$ } & $p_{T_{\widetilde{\mu},N}}$ & $0.9599$  \\  \cline{2-4}
 & & $\widetilde{\sigma}$ & $5.0011$ \\   \cline{3-4}
 &  & $\tilde{S}_N^{^\sigma}$ & $1.07$ $10^{-2}$ \\  \cline{3-4}
 & & MSE$_{\sigma}$ & $1.06$ $10^{-2}$\\  \cline{3-4}
\multirow{8}{*}{}&\multirow{4}{*}{$u_2=11=q_{_{81.17\%}}$ } &  $p_{T_{\widetilde{\sigma},N}}$ & $0.9909$   \\  \cline{2-4}
 & & $\widetilde{u_2}$ & $10.5898$  \\   \cline{3-4}
 &  & $\tilde{S}_N^{^{u_2}}$ & $1.567$  \\  \cline{3-4}
 &($u_1=4.16=q_{_{63.01\%}}$) & MSE$_{u_2}$ & $1.7196$\\  \cline{3-4}
\multirow{8}{*}{}&\multirow{4}{*}{$\xi=1.2$ } & $p_{T_{\widetilde{u_2},N}}$ & $0.7431$   \\  \cline{2-4}
 & & $\widetilde{\xi}$& $1.2312$ \\   \cline{3-4}
  &  & $\tilde{S}_N^{^\xi}$ & $8.88$ $10^{-3}$ \\  \cline{3-4}
 & & MSE$_{\xi}$ & $9.77$ $10^{-3}$ \\  \cline{3-4}
 & & $p_{T_{\widetilde{\xi},N}}$ & $0.7399$  \\   \hline
 \multicolumn{3}{|c|}{ Average  execution time (seconds)} &$16.31$  \\ \hline
  \multicolumn{3}{|c|}{ Average  iterations number} &$17$\\ \hline
 \multicolumn{3}{|c|}{ $\mathcal{D}$} &$9.28$ $10^{-4}$ \\ \hline
\end{tabular}
}
\end{table}
\end{minipage}

\end{document}